\documentstyle[11pt]{article}
\date{}

\begin{document}

\setcounter{page}{0}
\thispagestyle{empty}

\begin{flushright}
CERN-TH. 95-35
%hep-ph/9503221
\end{flushright}
\vspace{0.2cm}

\begin{center}
{\LARGE
{\bf REAL-TIME DESCRIPTION OF PARTON-HADRON CONVERSION
AND CONFINEMENT DYNAMICS}

}
\end{center}
\bigskip

\begin{center}

{\Large
{\bf John Ellis and Klaus Geiger}
}

{\it CERN TH-Division, CH-1211 Geneva 23, Switzerland}
\end{center}
\vspace{1.0cm}

\begin{center}
{\large {\bf Abstract}}
\end{center}
\smallskip

We propose a new and universal approach to the hadronization problem
that incorporates both partonic and hadronic degrees of freedom
in their respective domains of relevance, and that describes the conversion
between them within a kinetic field theory formulation in real time and full
7-dimensional phase space.
We construct a scale-dependent effective theory that reduces to
perturbative QCD with its scale
and chiral symmetry properties at short space-time distances, but
at large distances ($r\,\lower3pt\hbox{$\buildrel > \over\sim$}\,1$ fm)
yields symmetry breaking gluon and quark condensates plus
hadronic excitations.
The approach is applied to the evolution of  fragmenting $q\bar q$ and $gg$ jet
pairs
as the system evolves from the initial 2-jet configuration,
via parton showering and cluster formation, to the final yield of hadrons.
The phenomenological implications for $e^+e^-\rightarrow hadrons$  are
investigated,
such as the time scale of the transition, and its energy dependence, cluster
size and mass distributions.
We compare our results for particle production and Bose-Einstein correlations
with experimental data, and find an interesting possibility of extracting
the basic parameters of the space-time evolution of the system
from Bose enhancement measurements.
\noindent

\vspace{0.5cm}

%\leftline{PACS Indices: .....................}
\rightline{johne@cernvm.cern.ch}
\rightline{klaus@surya11.cern.ch}
\leftline{CERN-TH. 95-35, March 1995}

\newpage

\noindent {\bf 1. INTRODUCTION}
\bigskip

The physics of QCD exhibits different relevant excitations at
distinct length (or momentum) scales.
To give this notion a well-defined meaning, consider some
characteristic length scale $L_c$ of the order of 1 $fm$ that
crudely separates short- from long-distance physics.
At short space-time distances ($r\ll L_c$)
the relevant degrees of freedom are quarks and gluons,
effectively unconfined due to asymptotic freedom, and their
interactions are well described by perturbative QCD.
The theory exhibits chiral symmetry and (approximate) scale symmetry.
At large distances
($r \gg L_c$)
%($r\,\lower3pt\hbox{$\buildrel > \over\sim$}\, L_c$)
on the other hand, we are in the regime of hadronic degrees of freedom
and physical observable particles,
whose non-perturbative interactions are known to be described well by
chiral models.
In between these two regimes, in the range $r \approx L_c$,
our current knowledge is essentially limited to the
understanding that there must be a rather sudden dynamical establishment
of long-range order, i.e. some kind of ``phase transition''
from the unconfined, chiral- and scale-invariant phase of
partons to the hadronic phase with massive physical states
and broken symmetries.

The {\it dynamics} of this parton-hadron conversion and confinement
mechanism has scarcely been studied yet,
although QCD-inspired effective quark models that incorporate
confinement phenomenology in some way have been
exploited extensively to describe static hadron properties rather well
\cite{qmodels}.
This problem is particularly serious for attempts to describe the phenomenon
of hadronization in high-energy QCD processes.
The theoretical tools currently
available for studying QCD are inadequate to describe the
transformation from partonic to hadronic degrees of freedom as a dynamic
process:
perturbative techniques \cite{pQCD} are limited to the short-distance regime
where confinement
is not apparent, whilst effective low-energy chiral models \cite{chiral}
and QCD sum rules \cite{shifman79}, that incorporate confinement, lack partonic
degrees of
freedom.
On the other hand, common descriptions of parton
fragmentation \cite{webber86} are usually based on
ad hoc prescriptions to simulate hadron formation from parton decays.
In principle, lattice QCD \cite{karsch89} should be able to bridge the gap, but
in practice
dynamical calculations of parton-hadron conversion are not yet feasible.

The purpose of the present paper is to
give a detailed documenation of our progress towards a consistent, fully
dynamical
formulation of the non-static properties of
confinement, chiral symmetry breaking and hadron formation,
as recently proposed in Ref. \cite{ms38}. Aside
from the aforementioned arguments, these issues are of
great interest in the context of the QCD phase transition in the
early Universe when hadrons formed from unconfined quark-gluon matter,
or  in high-energy heavy-ion collisions, where
one expects a very hot and dense deconfined quark-gluon plasma
to be created.
Here it is inevitable to employ a dynamical treatment of the
transition from short-distance (perturbative) regime
of partons to the long-distance (non-perturbative) domain of hadrons.
We extend here previous work \cite{ms36} and
present a universal approach to the dynamic
transition {\it between} partons and hadrons based on an effective
QCD field theory description and relativistic kinetic theory.

Our concept is the following:
we start from a gauge-invariant Lagrangian formulation
that embodies both fundamental partonic and composite hadronic degrees of
freedom.
It is explicitly dependent on the space-time scale $L=\sqrt{r^2}$ at
which the physics is ``probed''.
The scale dependence  is however not external, but the variation of the scale
is governed by
the dynamical evolution of the physical system under consideration.
The field equations of motion can be cast into evolution equations for the
real-time Green functions of the various particle
species, and by following the space-time evolution
we can trace the conversion from
partonic to hadronic degrees of freedom in full 7-dimensional phase space,
as it is driven by the dynamics.
This  effective field theory approach
recovers QCD with its scale
and chiral symmetry properties at short distances or high momentum transfers,
but yields
at low energies the formation of symmetry breaking gluon and
quark condensates including excitations that represent
the physical hadrons.
\smallskip

It is important to explain in more detail
the physical basis of our approach:
we assume that the vacuum state in QCD can be visualized as a
``color dielectric medium'' \cite{leebook}
chracterized by some collective color-singlet fields
that correspond in the long-wavelength limit
to the gluon and quark condensates, and incorporate
phenomenologically the complex structure
of the physical vacuum as order parameters.
Specifically, the underlying hypothysis \cite{fai88,krein91} is that
the {\it long-distance} (non-perturbative) gluon self-interactions generate
an effective {\it scalar gluon condensate field} $\chi$ which is
self-interacting
through some potential $V$ constructed \cite{CEO} on the basis of the symmetry
properties of the
QCD Lagrangian. As a consequence of symmetry constraints,
the scalar field $\chi$ must in addition couple through the potential $V$ to an
effective {\it pseudoscalar quark condensate field} $U$, a
feature which is also  suggested by lattice QCD
studies \cite{karsch89}, indicating that the confining and chiral symmetry
breaking
``phase transitions'' are in some way related and occur approximately at the
same scale.
The long-range properties of the non-perturbative vacuum are then characterized
by gluon
and quark condensates which are proportional to the non-vanishing
vacuum expectation values  $\langle \chi \rangle \equiv \chi_0$ and
$\langle U \rangle \equiv U_0$ in the long-distance limit.

Our central idea is that the effective gluon condensate field $\chi$
plays the driving role in the generation of confinement and
chiral-symmetry-breaking mechanisms: the non-perturbative
gluon self-interactions are assumed to modify the long-range properties
of the vacuum in such a way that the propagation of the
{\it elementary gluon fields} $A^\mu$ is altered with increasing
space-time distances and eventually completely suppressed (confinement).
As a direct consequence \cite{krein91}, the self-energy of the
{\it elementary quark fields} $\psi$ will be modified accordingly through
the quark-gluon coupling, so that it generates dynamically an effective
quark mass term which becomes infinite at large space-time distances.
The coupling between the perturbative regime with elementary fields $A^\mu$,
$\psi$
and the non-perturbative vacuum represented by the condensate fields
$\chi$, $U$ is mediated by a single dimensionless
``color-dielectric function'' $\kappa(\chi)$ \cite{FL0}, which
vanishes in the short-distance limit, but approaches unity at large distances.
There is no need to introduce an additional coupling to
the field $U$, nor to consider an explicit chiral-symmetry-breaking
quark mass term, because, as mentioned,
the quarks aquire a dynamic mass via their coupling to the
gluons in the presence of the field $\chi$.

We thus obtain an effective QCD field theory which in the short
distance limit ($\langle \chi\rangle =0, \langle U\rangle =0, \kappa(0)=1$) is
chiral invariant
and incorporates free gluon and quark propagation (asymptotic freedom),
whereas in the long distance limit ($\langle \chi\rangle = \chi_0, \langle U
\rangle = U_0,
\kappa(\chi_0)=0$) no gluon or quark propagation can occur (confinement).
In between these two regimes, the effective theory interpolates and governs the
dynamics of the
conversion of short-distance fluctuations (partons) to non-perturbative
bound states (hadrons) embedded in the physical vacuum.
As a prototype case, we study in detail the parton-hadron conversion
in $e^+e^- \rightarrow hadrons$.
We visualize the process $e^+e^- \rightarrow \bar q \,q$
as producing a ``hot spot'' in which the long-range order
represented by $\chi_0$ and $U_0$ is disrupted locally by the
appearance of a bubble of the naive perturbative
vacuum.  Within this bubble, a parton shower develops
in the usual perturbative way, with the hot spot expanding
and cooling in an irregular stochastic manner described by QCD transport
equations.
This perturbative description remains appropriate in any
phase-space region of the shower where the local energy density is large
compared with the difference in energy density between
the perturbative partonic and the non-perturbative hadronic vacua.
When this condition is no longer satisfied, a bubble of hadronic
vacuum may be formed with a probability determined by statistical-mechanical
considerations.
\medskip

This paper is organized in two main parts. The first part,
consisting of Secs. 2 and 3, is intended as a comprehensive
presentation of  the general field-theoretical framework and
necessary elements of quantum transport theory.
In Sec. 2 we construct on the basis of  the dual vacuum picture of coexisting
perturbative and non-perturbative domains an effective theory
that embodies the correct scale and symmetry properties of QCD and
that has the desired features outlined above.
We also discuss the relation to the phenomenology of the QCD phase transition,
where the role of the critical temperature is analogous to the
critical confinement length scale in our approach.
Sec. 3 outlines the method of real-time Green functions that we use to
derive from the field equations of motion the corresponding  coupled
equations for the particle distribution functions.
We also indicate how macroscopic quantities related to  observables
can be extracted from the microscopic particle dynamics within the kinetic
theory
of (non-equilibrium) many-particle systems.
The second part of the paper, Secs. 4-6, is devoted to
the application of this effective QCD field theory
to the dynamics of parton-hadron conversion
for the prototype process of fragmenting jet systems
initiated by $e^+e^-$ annihilation.
In Sec. 4 we derive transport equations that, in the case of
the partons, are generalized QCD evolution equations in full phase space,
and similar equations for the excitations of the $\chi$ and $U$ fields.
In Sec. 5 we present results of simulating this
real-time evolution  of partons through the perturbative
shower stage, via subsequent formation of color-singlet clusters, and finally
hadronic cluster decay to give the final hadron yield.
We investigate phenomenological implications for particle
production and the Bose-Einstein effect, which we find to be a
particularly sensitive probe to measure and test the confinement dynamics.
Finally, Sec. 6 is reserved for a summary and a brief discussion of
future perspectives of the approach, in particular its applicability to
the QCD phase transition, and to high-density QCD.
\bigskip
\newpage

\noindent {\bf 2. FIELD THEORY FRAMEWORK}
\bigskip

As explained in the introduction, the vacuum state in QCD may be pictured as  a
color-dielectric medium characterized by long-range order parameters.
Consider the vacuum expectation values ($vev$'s) of the normal-ordered
products of color-singlet and Lorentz scalar (pseudoscalar)
functions $\chi$ ($U$) \cite{ms36,FL0}
\begin{equation}
\langle \,0 \,| \, \chi(F_a^{\mu\nu}) \, |\,0\, \rangle
\;\;\; ; \;\;\;\;\;\;\;\;\;\;\;\;\;
\langle \,0 \,| \, U(\psi_i, \overline{\psi}_j ) \, |\,0\, \rangle
\; ,
\label{vev1}
\end{equation}
where $F_a^{\mu\nu}$ is the usual $SU(3)$ field strength tensor,
$\psi$, $\overline{\psi}$ the quark fields, and $\chi(0)$, $U(0)$ is set
to be zero.
For instance, $\chi$ can be
$f_{abc}F_{\mu\nu}^a F_{\nu\lambda}^b F_{\lambda\mu}^c$
or $(F_{\mu\nu}^a F_{\mu\nu}^a)^2$, or other combinations.
Similarly, $U$ may be composed of $Tr[(\overline{\psi}_i \psi_j )^2]$,
$Tr[(\overline{\psi}_i \overline{\psi_j}\psi_k \psi_l )^2]$, etc..
These $vev$'s are physical quantities that characterize the structure
of the QCD vacuum \cite{Shuryakbook} and are related to the measurable gluon
and
quark condensates, respectively.

Clearly, if we take the long wavelength limit $L\rightarrow \infty$
and simultanously let the coupling strength $g_s$ among the fields tend to
zero, we have in general
the non-comutativity of the double limits
\begin{eqnarray}
\lim_{L\rightarrow \infty}
\lim_{g_s\rightarrow 0}\;
\langle \,0 \,| \, \chi(F_a^{\mu\nu}) \, |\,0\, \rangle
\;=\;0
& & \ne \;\;\;
\lim_{g_s\rightarrow 0}\;
\lim_{L\rightarrow \infty}
\langle \,0 \,| \, \chi(F_a^{\mu\nu}) \, |\,0\, \rangle
\nonumber
\\
\lim_{L\rightarrow \infty}
\lim_{g_s\rightarrow 0}\;
\langle \,0 \,| \, U(\psi_i, \overline{\psi}_j ) \, |\,0\, \rangle
\;=\;0
& & \ne \;\;\;
\lim_{g_s\rightarrow 0}\;
\lim_{L\rightarrow \infty}
\langle \,0 \,| \, U(\psi_i, \overline{\psi}_j ) \, |\,0\, \rangle
\;\;,
\label{vev2}
\end{eqnarray}
with $\langle 0 | F_a^{\mu\nu} |0 \rangle
= 0 =\langle 0 | \psi_i |0 \rangle$.
Eq. (\ref{vev2}) is a pure quantum phenomenon and a typical property of
phase transitions. It implies that there is long-range
order in the vacuum which can be characterized by the
operator functions $\chi$ and $U$.

In order to embody this concept into a field theory formulation,
let us define the  {\it distance measure} $L$ for
the space-time separation between two points $r$ and $r'$ ($r^\mu=(t,\vec r)$):
\begin{equation}
L\;\,:=\;\, \sqrt{(r-r')_\mu(r-r')^\mu}
\;,
\label{Ldef}
\end{equation}
and introduce a {\it characteristic length scale} $L_c$ that separates
short distance ($L\ll L_c$)
and long range ($L\,\lower3pt\hbox{$\buildrel > \over\sim$}\,L_c$)
physics in  QCD. The scale $L_c$ can be associated with the
confinement length of the order of a hadron radius,
as we will specify more precisely later.
\bigskip
\newpage

\noindent {\bf 2.1 The short-distance regime {\boldmath $L \ll L_c$}}
\medskip

At {\it small space-time distances} $L \ll L_c$, because of asymptotic freedom,
the properties of QCD are well described by a perturbative expansion in
powers of the coupling $g_s$ of the generating functional for the
connected Green functions,
\begin{equation}
W [ J,\eta,\overline{\eta}] \;=\;
\int {\cal D} A^\mu {\cal D} \psi {\cal D} \overline{\psi}\; \mbox{det} {\cal
F}
\;\exp\left\{
i\,\int d^4 r \left( {\cal L}[A^\mu,\psi,\overline{\psi}] \;+\;
J_{\mu , a} A^\mu_a \;+\;
\overline{\psi}_i \eta \;+\;\overline{\eta} \psi_i \right) \right\}
\;.
\label{genf1}
\end{equation}
In the path integral,
$\mbox{det} {\cal F}$ denotes the Fadeev-Popov determinant and
$J$, $\eta$, $\overline{\eta}$ are the generating currents for
the gluon fields $A_a^\mu$ and the  quark fields $\psi$, $\overline{\psi}$
(which are vectors in flavor space, $\psi\equiv (\psi_u,\psi_d,\ldots)$), and
the
QCD Lagrangian is
\begin{equation}
{\cal L}[A^\mu,\psi,\overline{\psi}] \;=\; -\frac{1}{4}\,\,F_{\mu\nu, a}
F^{\mu\nu}_a
\;+\;  \overline{\psi}_i \left[\frac{}{}\,(i \gamma_\mu \partial ^\mu
- m) \delta_{ij}
- g_s  \gamma_\mu A^\mu_a T_a^{ij} \right]\, \psi_j
\;+\;
\xi_a(A)
\label{LQCD}
\;,
\end{equation}
where
$F_a^{\mu\nu}= \partial^\mu A_a^\nu -\partial^\nu A_a^\mu + g_s f_{abc} A^\mu_b
A^\nu_c$
is the gluon field-strength tensor. The subscripts $a, b, c$ label the
color components of the gluon fields,
and $g_s$ denotes the color charge related to  $\alpha_s =g_s^2/(4\pi)$.
The $T_a$ are the generators of the $SU(3)$ color group, satisfying
$[T_a,T_b] = i f_{abc} T_c$ with the structure constants $f_{abc}$.
The indices $i,j$ label the color components of the quark fields and
$m\equiv \mbox{diag}(m_u,m_d,\ldots)$. Throughout, summation over the color
indices $a,b,c$ and $i,j$
is understood. We recall that
on setting the quark current masses $m$ to zero, one has exact chiral symmetry.
The gauge-fixing term is denoted by a general function
$\xi_a(A)$ which, e.g., in covariant gauges
is $\xi_a(A) \equiv -1/(2 \alpha) (\partial_\mu A_a^\mu)^2$
with Lagrange multiplier $1/\alpha$. However, we will later consider a
different
(ghost-free) gauge that is more convenient for our purposes.
\bigskip

\noindent {\bf 2.2 The long-distance regime
{\boldmath $L \gg L_c$}}
%{\boldmath $L \,\lower3pt\hbox{$\buildrel < \over\sim$}\, L_c$}}
\medskip

The long-range physics of QCD
at {\it large space-time distances}
$L \gg L_c$,
%$L \,\lower3pt\hbox{$\buildrel < \over\sim$}\, L_c$,
is known to be described well by an effective low-energy theory.
Here we adopt the approach of
Ref. \cite{CEO} and define the corresponding generating functional as:
\begin{equation}
W [ J_\chi, K_U, K_U^\dagger] \;=\;
\int {\cal D} \chi {\cal D} U {\cal D} U^\dagger\;
\;\exp\left\{
i\,\int d^4 r \left( {\cal L}[\chi,U,U^\dagger] \;+\;
J_\chi \chi \;+\;
U^\dagger K_U \;+\;K_U^\dagger U \right) \right\}
\;.
\label{genf2}
\end{equation}
The field degrees of freedom are
a {\it scalar gluon condensate field} $\chi$ and
a {\it pseudoscalar quark condensate field}
$U = f_\pi \exp\left(i \sum_{j=0}^8 \lambda_j\phi_j/f_\pi\right)$
for the nonet of the meson fields $\phi_j$
($f_\pi=93$ MeV, $Tr[\lambda_i\lambda_j]=2 \delta_{ij}$, $U
U^\dagger=f_\pi^2$),
with non-vanishing $vev$'s in the long-distance limit,
\begin{eqnarray}
\chi_0 &:=&
\frac{\delta}{\delta J_\chi}
\ln W [ J_\chi, K_U, K_U^\dagger] \;=\;
\frac{\langle\, 0\,|\;\chi\; |\,0\,\rangle}{\langle\, 0\,|\,0\,\rangle}
\; \ne \;0
\\
U_0 &:=&
\left(\frac{\delta}{\delta K_U^\dagger}
+ \frac{\delta}{\delta K_U}\right)
 \ln W [ J_\chi, K_U, K_U^\dagger] \;=\;
\frac{\langle \,0\,|\; U+U^\dagger\; |\,0\,\rangle}{\langle\, 0\,|\,0\,\rangle}
\; \ne \;0
\;,
\label{vev3}
\end{eqnarray}
and an effective action
\begin{eqnarray}
\Gamma[\chi,U,U^\dagger]
&\equiv&
\ln W [ J_\chi, K_U, K_U^\dagger] \;-\;
\,\int d^4 r \left\{
J_\chi \chi \;+\; U^\dagger K_U \;+\;K_U^\dagger U \right\}
\\
&=&
\int d^4 r \left\{
\;-\;V(\chi,U)
\;+\;\frac{1}{2}\,(\partial_\mu \chi) ( \partial^\mu \chi )
\;+\; \frac{1}{4}\,
Tr\left[\frac{}{}(\partial_\mu U)
( \partial^\mu U^\dagger )\right]
\;+\;\ldots \right\}
\;.
\nonumber
\label{effact}
\end{eqnarray}
Consequently the  Lagrangian in (\ref{genf2}) is given by
\begin{equation}
{\cal L}[\chi,U,U^\dagger] \;=\;
\frac{1}{2}\,(\partial_\mu \chi) ( \partial^\mu \chi )
\;+\; \frac{1}{4}\,
Tr\left[\frac{}{}(\partial_\mu U)
( \partial^\mu U^\dagger )
\right]
\;-\;V(\chi,U)
\;,
\label{Lchi}
\end{equation}
with a  potential $V$ that has been  constructed  \cite{schechter81,ellis84}
on the basis of constraints which
arise from the scale and chiral symmetry properties of the
excact QCD Lagrangian, namely,
\begin{eqnarray}
V(\chi, U) &=&
b \;\left[ \frac{1}{4} \,\chi_0^4 \;+\;
\chi^4 \,\ln\left(\frac{\chi}{e^{1/4} \chi_0}\right)\right]
\;+\;
\frac{1}{4}\,\left[1\,-\,\left(\frac{\chi}{\chi_0}\right)^2\right] \;
Tr \left[ (\partial_\mu U)(\partial^\mu U^\dagger)\right]
\nonumber \\
& & +\;
 c\; Tr\left[\frac{}{} \hat m_q (U + U^\dagger) \right]
 \; \left(\frac{\chi}{\chi_0}\right)^3
\;+\;
\frac{1}{2}\; m_0^2 \; \phi_0^2
\;\left(\frac{\chi}{\chi_0}\right)^4\;
\;.
\label{V}
\end{eqnarray}
Here the parameter $b$ is related to the conventional bag constant $B$ by
\begin{equation}
B\;=\;b\;\frac{\chi_0^4}{4}
\label{bag}
\;.
\end{equation}
Furthermore,
$c$ is a constant of mass dimension 3,  $m_q = \mbox{diag}(m_u,m_d,m_s)$
is the light quark mass matrix, and $m_0^2$ is an extra
U(1)-breaking mass term for the ninth pseudoscalar meson $\phi_0$
(which we will disregard in the following).
In the chiral limit, this potential has a  minimum when $\langle \chi\rangle
=\chi_0$
and equals the vacuum pressure  $B$ at $\langle \chi\rangle =0$.
\bigskip

\noindent {\bf 2.3 The intermediate regime {\boldmath $L \approx L_c$}}
\medskip

Having established a field theory framework for the two regions
$L \ll L_c$ and $L\gg L_c$,
the crucial issue is now the intermediate range.
Clearly there must be a  dynamical
interpolation around $L \approx L_c$ from the
short-range to the long-range description.
We propose here the following approach.
Let us first consider the long-range domain, i.e. the physical
vacuum characterized by $\chi_0$, and introduce into the
vacuum an excitation of small space-time extent.
For instance,
imagine the creation of a $q\bar q$ pair with invariant mass
$Q \simeq L^{-1} \gg L_c^{-1}$ by
a time-like virtual photon from $e^+e^-$ annihilation.
The insertion of such a localized excitation (``hot spot'') modifies the vacuum
and we assume that the corresponding change in the action integral
$S\equiv \int d^4r {\cal L}[\chi,U,U^\dagger]$ in (\ref{genf2})
can be evaluated sufficiently accurately to second order as
\begin{eqnarray}
\delta S &=&
\frac{1}{2} \,\int d^4 r \left\{
\left\langle\, 0\,\left|\;\frac{\delta^2 {\cal L}[\chi, U, U^\dagger]}
{\delta F_{\mu\nu, a} \delta F^{\mu\nu}_b}\; \right|\,0\,\right\rangle
\;F_{\mu\nu,a}F^{\mu\nu}_b \, \delta_{ab}\;+\;
\left\langle\, 0\,\left|\;\frac{\delta^2 {\cal L}[\chi, U, U^\dagger]}
{\delta \psi_i  \delta \overline{\psi}_j}\; \right|\,0\,\right\rangle
\psi_i  \overline{\psi}_j \,\delta_{ij}
\right\}
\nonumber
\\
&=&
\int d^4r \left\{ -\frac{\kappa_L(\chi)}{4}\,\,F_{\mu\nu, a} F^{\mu\nu}_a
\;-\;  \mu_L(\chi) \overline{\psi}_i\, \psi_i
\right\}
\;,
\label{Smod}
\end{eqnarray}
where $\kappa_L$ and $\mu_L$ refer to the appropriate $vev$'s. Note that this
change $\delta S$ in the action preserves local gauge invariance.
We also remark that this ansatz implicitly assumes that
the elementary gluon ($F_{\mu\nu}$) and quark fields ($\psi, \overline{\psi}$)
couple directly only to the scalar field $\chi$, but not to the pseudoscalar
field $U$. The dynamics of $U$ is solely driven by its
coupling to $\chi$ through the potential $V(\chi,U)$, eq. (\ref{V}).

On the other hand, we know that the short-range properties at $L\ll L_c$
of our $q\bar q$ excitation are not affected by the long-range correlations.
Thus, here we can use (\ref{LQCD}) with perturbative methods,
since the quanta are asymptotically free and
$\langle \chi \rangle = 0 = \langle U+U^\dagger \rangle$.
Thus we can combine (\ref{LQCD}) and the effect of (\ref{Smod})
by adding to
${\cal L}[A^\mu,\psi,\overline{\psi}]$ and ${\cal L}[\chi,U,U^\dagger]$
the following contribution that carries an explicit
scale- ($L$-)dependence:
\begin{equation}
{\cal L}_L[A^\mu,\psi,\overline{\psi},\chi] \;=\;
\int d^4r \left\{ \frac{1}{4}\left(\frac{}{}1-\kappa_L(\chi)\right)
\,\,F_{\mu\nu, a} F^{\mu\nu}_a
\;-\;  \mu_L(\chi)\,\overline{\psi}_i\, \psi_i
\;-\;\left(\frac{}{}1-\kappa_L(\chi)\right)\, \xi_a(A)
\right\}
\;,
\label{LQCDchi}
\end{equation}
where the third term in the integrand is necessary to maintain
local gauge invariance.
It remains to specify the form of the functions $\kappa_L(\chi)$ and
$\mu_L(\chi)$.
Since $\kappa_L$ has to satisfy the boundary conditions \cite{FL0}
\begin{equation}
\kappa_L(0)\;=\; 1
\;\;,\;\;\;\;\;\;\;\;\;\;\;\;\;\;
\kappa_L(\chi_0)\;=\; 0
\;,
\label{kappa1}
\end{equation}
and is constrained to be a Lorentz-invariant color-singlet function
of scale dimension zero, a minimal possibility is
\begin{equation}
\kappa_L(\chi)\;=\; 1\;-\; \left(\frac{L\,\chi}{L_0\,\chi_0}\right)^2
\;.
\label{kappa}
\end{equation}
It turns out that
the particular form of $\kappa_L(\chi)$ is not crucial as long as the
properties
(\ref{kappa1}) are satisfied \cite{wilets},
because parton-hadron conversion is quite rapid, as we see later,
being related to the weakly first-order nature of the QCD phase transition
at finite temperature.
The essence is that (\ref{kappa})
enforces color charge confinement due to the fact that
a color electric charge creates a displacement $\vec D_a = \kappa_L \vec E_a$,
where $E_a^k = F_a^{0k}$, with energy $\frac{1}{2} \int d^3 r D_a^2/\kappa_L$
which becomes infinite at large $r$ for non-zero total charge.

Similarly, absolute confinement can be ensured also for quarks by coupling
the quark fields to the $\chi$ field through
\begin{equation}
\mu_L(\chi) \;=\;  \mu_0\;\left(\frac{1}{\kappa_L(\chi)}\;-\;1\right)
\;=\; \frac{\mu_0\;(L\,\chi)^2}{(L_0 \chi_0)^2 - (L \chi)^2}
\;,
\label{Mchi}
\end{equation}
where $\mu_0$ is a constant of mass dimension one that we will set equal to 1
GeV.
This form
reflects that the quark mass term $\mu_L(\chi)$ in (\ref{LQCDchi}) is
induced by non-perturbative gluon interactions, rather than being an
independent quantity, as is suggested by an explicit calculation \cite{krein91}
of the quark self-energy involving the gluon propagator in the
presence of the collective field $\chi$.
It has been shown \cite{fai88} that the dynamical mass $\mu_L(\chi)$
leads to an effective confinement potential
with the masses of the quarks at small $L$
approximately equal to the current masses, but at large $L$ when
$\langle\chi\rangle \rightarrow\chi_0$ it
generates an infinite asymptotic quark mass,
\begin{equation}
\mu_L(0)\;=\; 0
\;\;,\;\;\;\;\;\;\;\;\;\;\;\;\;\;
\mu_L(\chi_0)\;=\; \infty
\;.
\label{Mchi1}
\end{equation}
It is evident from (\ref{kappa})-(\ref{Mchi1}) that
${\cal L}[A^\mu,\psi,\overline{\psi},\chi]$
given by (\ref{LQCDchi})
vanishes in the short-distance limit ($L\rightarrow 0$, $\langle \chi \rangle
\rightarrow 0$),
whereas in the long-distance limit it suppresses the propagation of colored
gluon and quark fluctuations, and interpolates smoothly between the two
extremes.
The typical functional forms of $\kappa_L(\chi)$ and $\mu_L(\chi)$ are
illustrated in Fig. 1.
\bigskip

\noindent {\bf 2.4 The scale-dependent generating functional for the effective
theory}
\medskip

Let us now summarize and combine the three contributions of Secs. 2.1-2.3
into a single action integral, and write down the resulting
generating functional as an effective description covering
the full range $0 < L < \infty$ and depending implicitly on the
scale $L$ as defined by (\ref{Ldef}):
\begin{eqnarray}
W_{L} [ J,\eta,\overline{\eta},J_\chi,K_U,K_U^\dagger] &=&
\int {\cal D} A^\mu  {\cal D} \psi {\cal D} \overline{\psi}
{\cal D} \chi  {\cal D} U {\cal D} U^\dagger
\; \mbox{det} {\cal F}
\label{genf3}
\\
& & \times \;
\;\exp\left\{
i\,\int d^4 r \left( {\cal L}[A^\mu,\psi,\overline{\psi}] \;+\;
{\cal L}_L[A^\mu,\psi,\overline{\psi},\chi] \;+\;
{\cal L}[\chi,U,U^\dagger]
\right.
\right.
\nonumber \\
& & \;\;\;\;\;\;\;\;\;
\left.
\left.
\frac{}{}
\;+\;J_{\mu , a} A^\mu_a \;+\;
\overline{\psi} \eta \;+\;\overline{\eta} \psi
J_\chi \chi \;+\;
U^\dagger K_U \;+\;K_U^\dagger U \right) \right\}
\;,
\nonumber
\end{eqnarray}
where
${\cal L}[A^\mu,\psi,\overline{\psi}]$ is given by (\ref{LQCD}),
${\cal L}[\chi,U,U^\dagger]$ by (\ref{Lchi}), and
${\cal L}_L[A^\mu,\psi,\overline{\psi},\chi]$ by (\ref{LQCDchi}), so that
the effective, $L$-dependent  Lagrangian density
${\cal L}_L \equiv {\cal L} [A^\mu,\psi,\overline{\psi}] +
{\cal L}[\chi,U,U^\dagger] +{\cal L}_L[A^\mu,\psi,\overline{\psi},\chi]$
in the generating
functional (\ref{genf3}) can be written as
\begin{eqnarray}
{\cal L}_L&=&
-\frac{\kappa_L(\chi)}{4}\,\,F_{\mu\nu, a} F^{\mu\nu}_a
\;+\;  \overline{\psi}_i \left[\frac{}{}\,\left(\frac{}{}i \gamma_\mu \partial
^\mu
- \mu_L(\chi)\right) \delta_{ij}
- g_s  \gamma_\mu A^\mu_a T_a^{ij} \right]\, \psi_j
\;+\;
\kappa_L(\chi)\,\xi_a(A)
\nonumber \\
& & +\;
\frac{1}{2}\,(\partial_\mu \chi) ( \partial^\mu \chi )
\;+\; \frac{1}{4}\,
Tr\left[\frac{}{}(\partial_\mu U)
( \partial^\mu U^\dagger )
\right]
\;-\;V(\chi,U)
\;,
\label{genf4}
\end{eqnarray}
where $V(\chi,U)$ is the potential given by (\ref{V}),
and we have gone over to the limit of zero current quark masses.
It is important to realize that the scale dependence of (\ref{genf3})
and (\ref{genf4}) arises solely through the
$L$-dependent functions $\kappa_L(\chi)$ and $\mu_L(\chi)$, given by
eqs. (\ref{kappa}) and (\ref{Mchi}).
The scale $L$ is {\it not} to be misunderstood as an external parameter.
Instead it is intrinsic variable of the formulation.
As we will see later, the variation
of $L$ is governed by the dynamics of the fields itself, and
it in turn determines the time evolution of the interacting
fields. Therefore, when studying the dynamical evolution of
some system under consideration, one must
necessarily require this self-consistency for a meaningful solution.
\medskip

\noindent

At this point let us state clearly the following important remarks:
\smallskip
\noindent

{\bf a)}
The effective field theory defined by (\ref{genf3}) and (\ref{genf4})
represents a description of the duality of
partonic and hadronic degrees of freedom: high-momentum,
short-distance quark-gluon fluctuations (the perturbative excitations) are
embedded
in a collective field $\chi$ (the non-perturbative vacuum), in which by
definition the low-momentum, long-range fluctuations are absorbed.
Confinement is thus associated with a dual stucture of the QCD vacuum.
The formulation is gauge- and Lorentz-invariant, and is consistent
with scale and chiral symmetry properties of QCD.
It interpolates between
the high-momentum (short-distance) QCD phase with unconfined gluon
and quark degrees of freedom and chiral symmetry
($\langle \chi\rangle=0, \langle U \rangle =0, \kappa_L =1, \mu_L=0$), to a
low-energy (long-range) QCD  phase with confinement and broken chiral
symmetry
($\langle \chi\rangle=\chi_0, \langle U \rangle =U_0, \kappa_L =0,
\mu_L=\infty$),
where $\chi_0$ and $U_0$ are the  long-range order parameters of
the vacuum, directly related to the gluon condensate  and
the quark condensate, respectively (Sec. 2.5 below).
\smallskip

\noindent
{\bf b)}
By introducing additional fields $\chi$ and $U$ to describe the long-range
behaviour of the gluon and quark fields, we must obviously be
careful not to double-count the degrees of freedom, since
the full theory of QCD is contained in
${\cal L} [A^\mu,\psi,\overline{\psi}]$, eq. (\ref{LQCD}),
already.
However, by our construction the  sum ${\cal L}_L$ in (\ref{genf4})
gives a consistent formulation that strictly avoids double counting, because
the introduction of the scale $L$ and the behaviour of the $L$-dependent
coupling functions $\kappa_L(\chi)$ and $\mu_L(\chi)$
truncate the dynamics of the elementary fields $A^\mu$, $\psi$
to the short-distance, high-momentum regime ($L \ll L_c$),
whereas the effective description
in terms of the collective fields $\chi$, $U$ covers the complementary
long-range, low-energy domain ($L \gg L_c$).
Accordingly, a quark or gluon is either considered a
colored short-range fluctuation (parton) or it is
part of a complex bound state (hadron), but not both.
\smallskip

\noindent
{\bf c)}
The presence of the non-linear coupling function $\kappa_L(\chi)$,
which also enters $\mu_L(\chi)$ via (\ref{Mchi}), means that
the sum $\Delta {\cal L}\equiv {\cal L}[\chi,U]+{\cal
L}[A^\mu,\psi,\overline{\psi},\chi]$
in  (\ref{genf4}) is non-renormalizable.
However, there
is no need for explicit renormalization, because
the composite fields  $\chi$ and $U$ are already interpreted as effective
degrees of freedom with loop corrections implicitly included in $\Delta {\cal
L}$,
and it would be double counting to add them again.
Moreover, as mentioned in item b)  above,
the low-energy domain of ${\cal L}[\chi,U]$
is by construction bounded from above by the onset of the high-energy regime
described by ${\cal L}[A^\mu,\psi,\overline{\psi}]$.
The  characteristic scale $L_c$ that separates the two
domains, therefore, provides an `ultra-violet' cut-off for
${\cal L}[\chi,U]$, and at the same time an `infra-red' cut-off for
${\cal L}[A^\mu,\psi,\overline{\psi}]$.
\bigskip

\noindent {\bf 2.5 Analogies with QCD at finite temperature}
\medskip

We close this Section with pointing out some
immediate phenomenological implications:
the particular form (\ref{V}) of the potential $V$
as a function of $\chi$, as well as the
functions $\kappa_L$ and $\mu_L$ that couple short- and long-range regimes,
play
a central role in dynamical processes where the scale $L$ changes with time.
The effect of (\ref{LQCDchi}) can be interpreted as a scale- ($L$-)dependent
modification $\delta V$, which adds to the ($L$-independent)
potential $V$, eq. (\ref{V}),
\begin{equation}
{\cal V}(L) \;:=\;
V(\chi, U) \;+\;\; \delta V(L,\chi)
\label{calV}
\;,
\end{equation}
\begin{equation}
\delta V(L,\chi) \;=\;
\frac{(L\chi)^2}{4\,(L_0\chi_0)^2} \; F_{\mu\nu, a} F^{\mu\nu}_a
\;+\;
\frac{\mu_0\;(L\chi)^2}{(L_0\chi_0)^2\,-\, (L \chi)^2} \;\overline{\psi}_i\,
\psi_i
\;.
\label{dV}
\end{equation}
where we used eqs. (\ref{kappa}) and (\ref{Mchi}).
We emphasize again that $L$ is not an input parameter, but rather
is determined by the space-time dependent separation of the colored quanta.
In Sec. 4, we will specify how to determine the variation of the variable $L$.

In view of (\ref{dV}), one has  $\delta V \propto O(L^{2})$,
therefore it is suggestive that the variable $L$ plays a similar role as the
temperature $T$
in finite-temperature QCD, where the correction to the zero-temperature
potential is $O(T^2)$ \cite{FTQCD}.
This formal analogy will be indicative in the following. However, one must bear
in
mind that here we are aiming to describe the evolution of a general
non-equilibrium
system in real time and Minkowski space, as opposed to thermal evolution
in Euclidean space.
Nevertheless, we adopt the general concept of
Ref. \cite{CEO}, and from the analogy with this previous work
we can qualitatively expect that the correction $\delta V$
will give a first-order ``phase transition'' from the parton to the hadron
phase,
when combined with $V$.

As seen in Fig. 2, there are three characteristic scales,
$L_\chi$, $L_c$ and $L_0$, that mark the time evolution from the small-$L$ to
the
large-$L$ region as the scale-dependent potential ${\cal V}(L)$, eq.
(\ref{calV}),
changes:
\smallskip

\noindent
{\bf (i)}
$L_\chi$ is the characteristic length scale below which the vacuum with
$\chi \ne 0$ cannot exist. The potential ${\cal V}$ has a unique
minimum at $\chi = 0$, i.e. we are in the perturbative vacuum of the pure
parton phase.
\smallskip

\noindent
{\bf (ii)}
$L_c$ marks the point when the ${\cal V}$ develops two degenerate
minima, one at $\chi=0$ and the other at $\chi=\chi_c$. The pressure in
the parton phase is here equal to the pressure in the hadron phase,
and the probability for partons to tunnel through the barrier becomes large.
\smallskip

\noindent
{\bf (iii)}
$L_0$ defines when $\delta V=0$ and ${\cal V}$ becomes equal to
$V$ in eq. (\ref{V}), and has a single absolute minimum
at $\langle \chi\rangle =\chi_0$. The parton phase cannot exist any longer, and
the
parton-hadron conversion is completed.
We are in the true (physical) vacuum characterized by the presence
of a gluon and a quark condensate.
\smallskip

Following \cite{CEO}, we can relate the
$vev$'s (\ref{vev3}), $\chi_0$ and $U_0$, to the gluon condensate
\begin{equation}
\langle\,0\,|\,\frac{\beta(\alpha_s)}{4 \alpha_s} \,F_{\mu\nu}F^{\mu\nu}
\,|\,0\,\rangle \;=\;-\,b \;\chi_0^4
\;\;\equiv\; G_0
\label{gcond}
\end{equation}
and the quark condensate
\begin{equation}
\langle\,0\,|\; \bar q q \;|\,0\,\rangle \;=\;
c\,\left(\frac{\chi}{\chi_0}\right)^3 \; U_0
\;\;\equiv\; Q_0
\label{qcond}
\;,
\end{equation}
\medskip
where $b$ and $c$ are defined in (\ref{V}). These condensates
can be regarded as local order parameters associated with gluon and
quark confinement, respectively,  and chiral symmetry breakdown.
Also, as discussed in \cite{CEO}, one can interpret  small oscillations about
the minimum of the potential ${\cal V} = V$ at $\langle \chi\rangle = \chi_0$,
$\langle U+U^\dagger\rangle =U_0$,
as physical hadronic states that emerge after symmetry breaking. They
include in principle:
(a) glueballs and hybrids as quantum fluctuations in the gluon condensate
$\chi_0$,
(b) pseudoscalar mesons as excitations of the quark condensate $U_0$,
(c) the pseudoscalar flavor singlet meson $\phi_0$, and
(d) baryons as topological solitons.
We will return below to the issue of hadron formation.
\bigskip

\noindent {\bf 3. EQUATIONS OF MOTION AND KINETIC EVOLUTION}
\bigskip

In this Section we outline how to obtain a fully dynamical description of
the parton-hadron conversion in real time and complete phase space,
starting from the defining generating functional (\ref{genf3}) of our effective
field theory.
A comprehensive derivation can be found in Ref. \cite{ms39}, to which we refer
for details.
The method is based on the Green function technique \cite{GFT1,GFT2}, here
applied  to
derive transport equations for the field operators $A^\mu$, $\psi$, $\chi$, and
$U$.
The form of the transport equations results directly from the
Dyson-Schwinger equations \cite{SDE}. The self-energy operators that
enter the connected part of the equations can then be evaluated
in a perturbative expansion. This leads to corresponding equations of
motion for the distribution functions of  particles, namely
gluons and quarks as colored fluctuations, and  scalar and pseudoscalar
hadronic excitations.
The solution for the time development of these particle distribution
functions in phase space will then
allow us to calculate macroscopic observable quantities within
the framework of relativistic kinetic theory.
\bigskip

\noindent {\bf 3.1 Equations of motion}
\medskip

We start from the field equations of motion that follow from
the variation of the generating functional (\ref{genf3}) with (\ref{genf4}):
\begin{eqnarray}
& &
\left[ \left(\frac{}{} i \gamma_\mu \partial^\mu
\,-\,\mu_L(\chi) \right) \delta_{ij}
\,-\, g_s \gamma_\mu A^\mu_a T_a^{ij} \right] \;\psi_j \;=\;0
\label{eom1}
\nonumber
\\
& &
\partial_\mu  \, F^{\mu\nu}_a\,+\,
g_s \, f_{abc} \, A_{\mu, b} F^{\mu\nu}_c
\,-\,
\left(\frac{}{}\partial_\mu \ln\kappa_L (\chi)\right)\, F^{\mu\nu}_a
\,-\,
\left(1+\frac{\mu_L(\chi)}{\mu_0}\right)\,g_s \,\overline{\psi}_i \,\gamma^\nu
T_a^{ij} \,\psi_j
\,+\,\xi^\nu_a(A)
\;=\;0
\label{eom2}
\nonumber
\\
& &
\partial_\mu \partial^\mu \chi \;+\;
\frac{\partial V(\chi,U)}{\partial \chi}\;+\; \frac{1}{4}\frac{\partial
\kappa_L(\chi)}{\partial \chi}\,
F_{\mu\nu, a} F^{\mu\nu}_a \;+\;
\frac{\partial \mu_L(\chi)}{\partial \chi}\, \overline{\psi}_i \psi_i \;=\;0
\label{eom3}
\nonumber
\\
& &
\partial_\mu \partial^\mu U \;+\;
\frac{\partial V(\chi,U)}{\partial U}
\;+\;\partial_\mu \frac{\partial V(\chi,U)}{\partial (\partial_\mu U)}\;=\;0
\label{eom4}
\;.
\end{eqnarray}
where $\partial^\mu\equiv\partial/\partial x^\mu$, and we set
$U\equiv (U+U^\dagger)/2$.
In the second equation, the function
\begin{equation}
\xi^\nu_a(A)\;:=\; \partial_\mu \left(\kappa_L\,\frac{\partial
\xi_a(A)}{\partial(\partial_\mu A_a^\nu)}\right)
\end{equation}
results from the gauge-fixing constraint in eq. (\ref{LQCD}), e.g., in
covariant gauges
$\xi^\nu_a(A)= (1/\alpha) \partial^\nu \partial_\mu A^\mu_a$.
Note that in general there are additional equations of motion involving
the ghost fields coupled to the gluon fields,
however, since we will later choose a ghost free gauge, these decouple and are
irrelevant here.

It is evident  from the above  equations and
the form of the potential $V(\chi,U)$, eq. (\ref{V}), that in the
short-distance
limit when $\langle \chi\rangle = 0$  and
$\kappa_L(\chi)=1$, the system of equations
decouples and reduces to the usual Yang-Mills equations. Similarly, in the
long-wavelength limit, one has $\langle \chi\rangle \rightarrow\chi_0$,
$\langle U\rangle \rightarrow U_0$, and $\kappa_L(\chi)\rightarrow 0$,
so that the dynamics in this case is completely described by the
equations for the effective fields $\chi$ and $U$.
A very important point is
that the $U$ field does not couple directly to the
quark or gluon fields. By construction \cite{CEO}, the dynamics of
the quark condensate field $U$ is solely driven by the
gluon condensate field $\chi$.
As a consequence,  the equation for $U$ is readily solved, once the
solution for $\chi$ is known.
It is important to realize that
the interplay between the $\chi$ field and the quark and gluon fields,
$\psi$ and $A$, is the crucial element of this approach.
{}From a phenomenological point of view, this implies that
the transformation of parton to hadron degrees of freedom proceeds
first by formation of scalar color-singlet states which subsequently
decay into pseudoscalar excitations (Sec. 4).
\bigskip

\noindent {\bf 3.2 Real-time Green functions and microscopic kinetics}
\medskip

The central role in the following is played by the {\it real-time Green
functions}
in the `closed-time-path formalism'  (for an extensive  review, see
\cite{GFT2}).
This formalism is the appropriate tool to describe general non-equilibrium
systems,
and its particular strength lies in the possibility of studying the time
evolution
of phenomena where initial and final states correspond to different vacua, as
we
are addressing here. In \cite{ms39} it is shown how one obtains a dynamical
formulation
which systematically incorporates quantum correlations and describes naturally
the transition from the perturbative QCD regime to the non-perturbative
QCD vacuum. The  real-time Green functions are
defined as the two-point functions that measure the {\it time-ordered
correlations}
between the fields at space-time points $x$ and $y$ (as before we suppress the
spinor indices for the fermion operators):
\begin{eqnarray}
& &
i\,S_{ij}(x,y)\;:=\; \langle \;T\, \psi_i(x)\, \overline{\psi}_j(y) \;\rangle
\label{green1}
\nonumber
\\
& &
i\,D_{ab}^{\mu\nu}(x,y)\;:=\; \langle \;T\, A^\mu_a(x)\, A^\nu_b(y) \;\rangle
\label{green2}
\nonumber
\\
& &
i\,\Delta(x,y)\;:=\; \langle \;T\, \chi(x)\, \chi(y) \;\rangle
\;-\;\langle\; \chi(x) \;\rangle \;\langle\; \chi(y) \;\rangle
\;,
\label{green3}
\nonumber
\\
& &
i\,\tilde{\Delta}(x,y)\;:=\; \langle \;T\, U(x)\, U^\dagger(y) \;\rangle
\;-\;\langle\; U(x) \;\rangle \;\langle\; U^\dagger (y) \;\rangle
\;,
\label{green4}
\end{eqnarray}
where $\langle \ldots \rangle$ denotes the vacuum expectation value, or,
in a  medium, the appropriate ensemble average, and
$T\,A(x)B(y) = \theta (x_0,y_0) A(x) B(y) \pm \theta(y_0,x_0) B(y) A(x)$
stands for the generalized time-ordering operator along a closed time contour
\cite{GFT2}
with $+$($-$) for boson (fermion) operators.
For the $\chi$ and $U$ fields, we have subtracted the classical expectation
values in order to
separate the quantum fluctuations of the fields around the classical field
configuration.
These Green functions generally describe the propagation of an excitation in a
many-particle system
from space-time point $x$ to point $y$. In the absence of a coupling
between $\psi$, $A^\mu$ and $\chi$, $U$,
and  in the zero-density limit, the expectation values $S_{ij}$ and
$D_{ab}^{\mu\nu}$ can be shown \cite{ms39} to reduce to the
usual quark and gluon Feynman propagators, respectively, and $\Delta$ with
the scalar Feynman propagator.

Using the definitions (\ref{green4}) and implementing the gauge-fixing
constraint
in (\ref{eom2}) for the gluon fields, one  finds
from (\ref{eom4})
the following equations of motion for the Green functions
(in the limit of zero rest masses),
\begin{eqnarray}
 i \gamma\cdot \partial_x\;S_{ij}(x,y)
&=&
\delta_{ij} \delta^4(x,y)\;+\;
\int d^4x' \,\Sigma_{ik}(x,x')\,S_{kj}(x',y)
\label{eog1}
\nonumber
\\
\partial_x^2 \;D^{\mu\nu}_{ab}(x,y)
&=&
\delta_{ab} \delta^4(x,y)\,\left(\frac{}{} g^{\mu\nu} \,-\,{\cal
E}^{\mu\nu}\right)
\;-\;\int d^4x' \,\Pi^\mu_{\sigma ,\;a,b'}(x,x')\,D^{\sigma \nu}_{b'b}(x',y)
\label{eog2}
\nonumber
\\
\partial_x^2\;\Delta(x,y)
&=&
- \delta^4(x,y)\;+\;
\int d^4x' \,\Xi(x,x')\,\Delta(x',y)
\label{eog3}
\nonumber
\\
\partial_x^2\;\tilde{\Delta}(x,y)
&=&
- \delta^4(x,y)\;+\;
\int d^4x' \,\tilde{\Xi}(x,x')\,\tilde{\Delta}(x',y)
\label{eog4}
\;,
\end{eqnarray}
describing the change with respect to $x$,
plus similar equations for the change with $y$ by the
substitions  $\partial_x \rightarrow -\partial_y$,
and $\Sigma(x,x')S(x',y) \rightarrow S(x,x')\Sigma(x',y)$, etc..
The explicit expressions for the {\it self energies} $\Sigma$, $\Pi$, $\Xi$,
and $\tilde{\Xi}$
are rather lengthy and can be found in Ref. \cite{ms39}.
We remark that the  functions  $\Sigma$ and $\Pi$
include both the usual quark and gluon self energies, as well as the
additional interaction of the quanta with the $\chi$ field.
Similarly, the self energy $\Xi$  incorporates the
effective self interaction of the $\chi$ field described by
the potential (\ref{V}), plus the interaction with the quark and gluon fields
in (\ref{dV}). Finally,
the self energy of the $U$ field plus its coupling to the $\chi$ field is
described by $\tilde{\Xi}$.
In the equation for the gluon Green function $D^{\mu\nu}_{ab}$,
we have on the right-hand side the remnant of the gauge constraint,
\begin{equation}
{\cal E}_{\mu\nu}\;:=\;\int \frac{d^4 p}{(2\pi)^4} \,\frac{e^{-i \,p\cdot
(x-y)}}{p^2+i0^+} \;\left[ g_{\mu\nu}
\,+\,\sum_{s=1,2} \,\varepsilon_\mu^\lambda (p,s)\,
\varepsilon_{\lambda\nu}^\ast(p,s)
\right]
\label{polsum}
\end{equation}
which involves a sum over the two physical (transverse) gluon polarizations
(e.g. in Feynman gauge
$\varepsilon_\mu^\lambda = g_\mu^\lambda$ and thus
$\sum_s \varepsilon_\mu^\lambda \varepsilon_{\lambda\nu}^\ast= -g_{\mu\nu}$,
i.e.
${\cal E}_{\mu\nu} = 0$).
We note that equations (\ref{eog4}) are of the form of
Dyson-Schwinger equations \cite{SDE}, and can be rewritten in symbolic operator
notations as
\begin{eqnarray}
S \;=\; S_0 \;+\; S_0\,\Sigma\, S
\;\;,\;\;\;\;\;\;\;\;
& &
D \;=\; D_0 \;-\; D_0\,\Pi\, D
\nonumber \\
\Delta \;=\; \Delta_0 \;+\; \Delta_0\,\Xi\, \Delta
\;\;,\;\;\;\;\;\;\;\;
& &
\tilde{\Delta} \;=\; \tilde{\Delta}_0 \;+\; \tilde{\Delta}_0\,\tilde{\Xi}\,
\tilde{\Delta}
\label{sde}
\;\;,
\end{eqnarray}
where $S_0,\;D_0,\;\Delta_0,\;\tilde{\Delta}_0$ denote the free-field Green
functions that
satisfy the equations of motion in the absence of self and mutual interactions.
Fig. 3 illustrates the diagrammatic representation of the Green functions
$S, D, \Delta, \tilde \Delta$, the self-energies $\Sigma, \Pi, \Xi, \tilde
\Xi$,
and the Dyson-Schwinger equations (\ref{sde}).

A quantum transport formalism can be derived
from the equations (\ref{eog4})
that is very suitable for the present purposes \cite{ms39}.
We confine ourselves here to sketching the essential steps.
One introduces the  {\it Wigner transforms} ${\cal W}$ \cite{wigner}
of the Green functions and the self energies
$W\equiv S, D,\Delta,\Sigma,\Pi,\Xi$:
\begin{equation}
{\cal W}(r,p) \;=\;
\int d^4 R \, e^{ip\cdot R}\; W\left( r,R\right)
\;\;,
\label{W}
\end{equation}
where
\begin{equation}
W(r,R)\; \equiv\; W\left(r+\frac{R}{2}, r-\frac{R}{2}\right)
\;=\; W(x,y)
\;,
\end{equation}
with $r\equiv (x+y)/2$ and $R\equiv x-y$
denoting  the center-of-mass and relative coordinates, respectively,
and $R$ being the canonical conjugate to the momentum $p$
(as before $r$, $p$, etc., denote four vectors, and $a\cdot b \equiv a_\mu
b^\mu$).
The  equations of motion for the Wigner transforms ${\cal W}(r,p)$
are now obtained \cite{ms39} under the assumption
that the Green functions and self-energies $W(r,R)$
can be approximated by a gradient expansion in $r$ up to first
order:
\begin{equation}
W(r+R, R) \;\simeq \; W(r, R) \;+ \;R \,\cdot\, \frac{\partial}{\partial
r}\,W(r,R)
\label{gradexp}
\;.
\end{equation}
This assumption implies a restriction to quasi-homogenous or moderately
inhomogenous
systems, such that the Green functions vary only slowly with $r$.
In homogenous systems, such as the vacuum, translation invariance dictates that
the dependence
on $r$ drops out entirely, and the Wigner transforms then coincide with
the momentum-space Fourier transforms of the Green functions and self energies.
Although we will in the present paper consider the evolution of a
fragmenting jet system in vacuum, the subsequent formulation
is tailored to apply also to more general translation {\it non}-invariant
situations in  moderately inhomogenous media.

The Dyson-Schwinger equations (\ref{eog4}) can now be converted
into kinetic  equations
by performing the Wigner transformation (\ref{W})
for all Green functions and self energies, and using (\ref{gradexp}).
One arrives at two
distinct equations for each of the Wigner transforms ${\cal S}$, ${\cal D}$ and
$\Delta$ ($\tilde{\Delta}$), with rather different physical interpretations:
(i) a {\it transport equation}, and (ii) a {\it constraint equation}.
The {\it transport equations} are
\begin{eqnarray}
& &
p\cdot \partial_r\;  {\cal S}_{ij} (r,p)
\;=\;
\frac{1}{2} \,\delta_{ij}\,\gamma\cdot \partial_r \;+\; \frac{1}{4}
(\gamma\cdot p\,+\, \Sigma)
 \;{\cal F}_{ij}^{(+)}
 \;+\;\frac{i}{8} \gamma\cdot \partial_r \;{\cal F}_{ij}^{(-)}
\nonumber
\\
& &
p\cdot\partial_r\; {\cal D}^{\mu\nu}_{ab} (r,p)
\;=\;
\frac{1}{4}\;{\cal G}^{\mu\nu\,(+)}_{ab}
\nonumber
\\
& &
p\cdot\partial_r\;\Delta(r,p)
\;=\;
-\frac{1}{4}\;{\cal H}^{(+)}
\label{T}
\;,
\end{eqnarray}
whilst the {\it constraint equations} are
\begin{eqnarray}
& &
\left[
\left(p^2 -\frac{1}{4} \partial_r^2\right) \delta_{ik}
\,-\,\Sigma^2_{ik}(r,p) \right]
\; {\cal S}_{kj} (r,p)
\;=\;
\delta_{ij}\,(\gamma\cdot p+ \Sigma) \,+\,
\frac{i}{4} (\gamma\cdot p+\Sigma) \,{\cal F}_{ij}^{(-)}
\,-\,\frac{1}{8}\,\gamma \cdot \partial_r \,{\cal F}_{ij}^{(+)}
\nonumber
\\
& &
\left[
g^\mu_\sigma \left(p^2 \,-\,\frac{1}{4} \partial_r^2 \right)\,\delta_{ac}
\;-\; \Pi^\mu_{\sigma,\, a,c}(r,p) \right]
\; {\cal D}^{\sigma\nu}_{cb} (r,p)
\;=\;
-\,\delta_{ab}\,\left(\frac{}{} g_{\mu\nu} -{\cal E}^{\mu\nu}\right) \;+\;
\frac{i}{4}\;{\cal G}^{\mu\nu\, (-)}_{ab}
\nonumber
\\
& &
\left[
p^2 \,-\,\frac{1}{4} \partial_r^2 \;+\; \Xi(r,p) \right]
\; \Delta (r,p)
\;=\;
1\;-\; \frac{i}{4}\;{\cal H}^-
\;.
\label{R}
\end{eqnarray}
The equations for $\tilde{\Delta}$ are formally identical to
those of $\Delta$.
The operator functions ${\cal F}$, ${\cal G}$, and ${\cal H}$
($\tilde{{\cal H}}$), which  include the effects of spatial inhomogenities, are
given by
($\partial_r^\mu\equiv \partial/\partial r^\mu$
$\partial_p^\mu\equiv \partial/\partial p^\mu$):
\begin{eqnarray}
{\cal F}_{ij}^{(-)}&=&
\left(
\left[ \partial^\mu_p \Sigma \, , \partial_\mu^r {\cal S}\right]_-
-
\left[ \partial^\mu_r \Sigma \, , \partial_\mu^p {\cal S}\right]_-\right)_{ij}
\;\;\;\;\;
{\cal F}_{ij}^{(+)}\;=\;
\left(
\left\{ \partial^\mu_p \Sigma \, , \partial_\mu^r {\cal S}\right\}_+
-
\left\{ \partial^\mu_r \Sigma \, , \partial_\mu^p {\cal
S}\right\}_+\right)_{ij}
\nonumber
\\
{\cal G}^{\mu\nu\,(-)}_{ab}&=&
\left(\left[ \partial^\mu_p \Pi \, , \partial_\mu^r {\cal D}\right]_-
-
\left[ \partial^\mu_r \Pi \, , \partial_\mu^p {\cal D}\right]_-
\right)^{\mu\nu}_{ab}
\;\;\;\;\;
{\cal G}^{\mu\nu\,(+)}_{ab}\;=\;
\left(\left\{ \partial^\mu_p \Pi \, , \partial_\mu^r {\cal D}\right\}_+
-
\left\{ \partial^\mu_r \Pi \, , \partial_\mu^p {\cal D}\right\}_+
\right)^{\mu\nu}_{ab}
\nonumber
\\
{\cal H}^-&=&
\;\;
\left[ \partial^\mu_p \Xi \, , \partial_\mu^r \Delta\right]_-
\;-\;
\left[ \partial^\mu_r \Xi \, , \partial_\mu^p \Delta\right]_-
\;\;\;\;\;\;\;\;\;
{\cal H}^+\;=\;
\left\{ \partial^\mu_p \Xi \, , \partial_\mu^r \Delta\right\}_+
\;-\;
\left\{ \partial^\mu_r \Xi \, , \partial_\mu^p \Delta\right\}_+
\;.
\end{eqnarray}

Eqs. (\ref{T}) and (\ref{R}) are our general master equations.
The physical significance \cite{ms39} of the transport equations (\ref{T})
and the constraint equations (\ref{R})
is that the former essentially describe the space-time
evolution of the Wigner transforms, whereas the constraint
equations describe the ``orthogonal'' evolution in momentum space,
and express a normalization condition imposed by unitarity
and the renormalization group.
In order to relate these operator equations to physically-relevant (observable)
quantities,
we define the {\it Wigner operators} $\hat F_\alpha(r,p)$
($\alpha\equiv q, g, \chi, U$)
in terms of the operators ${\cal S}$, ${\cal D}$, $\Delta$, $\tilde{\Delta}$
and the self energies $\Sigma$, $\Pi$, $\Delta$, $\tilde \Delta$
as follows:
\begin{eqnarray}
i {\cal S}_{ij}(r,p)&=&
\delta_{ij} \,\left(\gamma\cdot p + \Sigma\right)
\;(2\pi i) \,\delta\left(p^2 - \Sigma^2 \right)\;\, \hat F_q(r,p)
\nonumber
\\
i {\cal D}^{\mu\nu}_{ab}(r,p)&=&
\delta_{ab} \;\varepsilon^{\mu\sigma}(p,s)\varepsilon_\sigma^{\nu\,\ast}(p,s)
\;(2\pi i) \,\delta\left(p^2 - \Pi \right)\;\, \hat F_g(r,p)
\nonumber
\\
i \Delta(r,p)&=&
(2\pi i) \,\delta\left(p^2 -  \Xi \right)\;\, \hat F_\chi(r,p)
\nonumber
\\
i \tilde{\Delta}(r,p)&=&
(2\pi i) \,\delta\left(p^2 - \tilde{\Xi} \right)\;\, \hat F_U(r,p)
\label{WW}
\;.
\end{eqnarray}
Then, by tracing over color and spin polarizations, and taking
the expectation values (or, in a medium, the
ensemble average) of these Wigner operators,
one obtains the scalar functions
\begin{equation}
 F_\alpha(r,p)\;\,\equiv\; \, F_\alpha (t, \vec r; \vec p, p^2=M_\alpha^2)
\;\;\;\;\;\;\;\;\;\;\;\; (\,\alpha\; = \;q,\; g,\; \chi,\; U\,)
\label{WWW}
\end{equation}
with
\begin{eqnarray}
 F_q(r,p)\;=\; \,\langle \; Tr[ {\cal S}(r,p) ] \;\rangle
& & ,\;\;\;\;\;\;\;\;\;\;\;\;
M_q^2 \;=\; \Sigma^2(r,p)
\nonumber \\
 F_g(r,p)\;=\; \,\langle \; Tr[ {\cal D}(r,p) ] \;\rangle
& & ,\;\;\;\;\;\;\;\;\;\;\;\;
M_g^2 \;=\; \Pi(r,p)
\nonumber \\
 F_\chi(r,p)\;=\; \,\langle \; \Delta(r,p) \;\rangle
\;\;\;
& & ,\;\;\;\;\;\;\;\;\;\;\;\;
M_\chi^2 \;=\; \Xi(r,p)
\nonumber \\
 F_U(r,p)\;=\; \,\langle \; Tr[ \tilde{\Delta}(r,p) ] \;\rangle
& & ,\;\;\;\;\;\;\;\;\;\;\;\;
M_U^2 \;=\; \tilde{\Xi}(r,p)
\label {WWWW}
\;.
\end{eqnarray}
The $c$-number functions $F_\alpha(r,p)$
are the quantum-mechanical analogues of the
classical phase-space distributions that measure the number of particles
at time $t$ in a 7-dimensional phase-space element $d^3rd^4p$.
Due to the effects of the self energies, three-momentum and energy are
generally independent variables, because the quanta can be off mass shell,
i.e., for zero rest masses,
$E^2=\vec p^{\,2} + M_\alpha^2 \ne \vec p^{\,2}$, where $M_\alpha^2$,
eq. (\ref{WWWW}), represents the off-shellness
due to the self and mutual interactions of the quanta ($M^2=0$ for on-shell
particles).
In contrast to the classical  propagation of on-shell particles, the Wigner
functions
(\ref{WW}) incorporate the quantum ``Zitterbewegung'' even in the absence of
interactions with other particles.
These spatial fluctuations arise from the combination
$p^2 - \partial^2_r/4$ acting on the Wigner operators in (\ref{R}),
and account for the uncertainty in spatial localization of a quantum particle
due to
its momentum that is determined by the particle's self interaction.
The self-consistent solution of the transport and constraint
equations (\ref{T}) and (\ref{R}) therefore corresponds to summing over
all possible quantum paths $r$ in space-time with fluctuations in
energy-momentum $p$,
constrained by the uncertainty principle \cite{ms39}.
This simultanous evolution in $r$ and $p$ of the Wigner functions $F_\alpha$ is
illustrated in Fig. 4.
\bigskip

\noindent {\bf 3.3 Macroscopic quantities}
\medskip

The functions $F_\alpha$, eq. (\ref{WWW}),
contain the microscopic information that is required for the
statistical description of the time evolution of a many-particle system in
complete phase space \cite{msrep}.
Depending on the physical situation under consideration, one
starts from specified initial distributions at $t=t_0$ and follows the
time evolution of the phase-space densities
$F_\alpha(r,p)$ according to the master equations (\ref{T}) and (\ref{R}).
At any time $t > t_0$, $F_\alpha(r,p)$
reflects the state of the system around $\vec r$ and $\vec p$.
One can then calculate, in a Lorentz-invariant manner,
directly from the microscopic densities $F_\alpha$,  the relevant macroscopic
quantities that are related to  observables.
Relativistic transport theory \cite{degroot} relates
physical quantities  to phase-space integrals over products of combinations of
four-momenta or tensors and the particle distributions.
Specifically, using (\ref{WW}), performing the traces over
color and spin indices where necessary, and taking the ensemble average, one
obtains
the local space-time-dependent particle currents $n_\alpha$
and the corresponding energy-momentum tensors $T_\alpha^{\mu\nu}$
for the different particle species $\alpha = q,\bar q, g, \chi, U$,
which are given by \cite{msrep}
\begin{equation}
n_\alpha^\mu (r) \;=\;
\int d\Omega_\alpha \, p^\mu F_\alpha (p,r)
\;,\;\;\;\;\;\;\;
T_\alpha^{\mu \nu} (r)
\;=\;
\int d\Omega_\alpha p^\mu p^\nu \, F_\alpha (r, p)
\label{Tmunu}
\;,
\end{equation}
where
$d\Omega_\alpha=\gamma_\alpha  dM^2 d^3 p/(16\pi^3p^0)$,
the $\gamma_\alpha$ are degeneracy factors for the internal
degrees of freedom (color, spin, etc.),
$M_\alpha$ measures the amount by which a particle $\alpha$ is off mass shell
as a result of the self energy terms in (\ref{T}), (\ref{R}),
and $p^0\equiv E= +\sqrt{\vec p^{\,2} + M_\alpha^2}$.
These macroscopic quantities can be written in  Lorentz-invariant form
by introducing for each species $\alpha$ the associated  matter flow velocity
$u_\alpha^\mu(r)$,
defined as a unit-norm time-like vector at each space-time point,
$(u_\mu u^\mu)_\alpha = 1$. A natural choice is
e.g. $u_\alpha^\mu=n_\alpha^\mu /\sqrt{ n_{\nu\,\alpha} n_\alpha^\nu}$.
By contracting the quantities (\ref{Tmunu}) with the
local flow velocities $u_\alpha^\mu$, one can now obtain
corresponding invariant scalars of  particle density,
pressure, and energy density, for each particle species $\alpha$ individually.
For instance, in the absence of viscosity:
\begin{eqnarray}
n_\alpha (r) & = & n_{\mu\,\alpha} (r) \;u_\alpha^\mu (r)
\label{nr}
\nonumber
\\
P_\alpha (r) & = & - \frac{1}{3} \,T_{\mu \nu ,\,\alpha} (r) \,
\left( \frac{}{} g^{\mu \nu}- \; u_\alpha^\mu (r) u_\alpha^\nu (r) \right)
\label{pr}
\\
\varepsilon_\alpha (r) & = & T_{\mu \nu ,\,\alpha} (r)  \; u_\alpha^\mu (r)
u_\alpha^\nu (r)
\label{er}
\nonumber
\;\;\;.
\end{eqnarray}
Due to the scalar character of these quantities, they provide
local, Lorentz-invariant measures of the many-particle system.
It is left to convenience in which Lorentz frame
the calculation is performed, but in general
it is considerably simplified in the local rest frame of the matter
where $u^\mu = ( 1, \vec 0)$.
The total number and the free energy of the particles
at a given time can then be obtained by integrating over position space.
\bigskip
\medskip

\noindent {\bf 4. PARTON-HADRON CONVERSION OF FRAGMENTING JET SYSTEMS}
\medskip

The preceding kinetic formulation allows us now to apply
the conceptual ideas of the effective field theory of Sec. 2 to
the dynamics of parton-hadron conversion in rather general situations.
In accord with the  formalism of Sec. 2, the parton-hadron
transition can be visualized as the conversion of high-momentum colored
quanta of the fundamental quark and gluon fields into color-neutral
composite states that correspond to local excitations of
the condensate fields $\chi$ and $U$ embedded in the physical vacuum.

Ultimately, we would like to address the dynamics of the
(non-equilibrium) QCD phase transition in finite-temperature
systems.
Here, however, we will as a first application study a
much simpler sytem, namely
the fragmentation of a $q\bar q$ jet system
with its emitted bremsstrahlung gluons, and describe the evolution of the
system as it converts from the parton phase to the hadronic phase (illustrated
in Fig. 5).
A time-like virtual photon in an $e^+e^-$ annihilation event
with large invariant mass $Q \gg \Lambda_{QCD}$ ($\Lambda_{QCD}=200 - 400$
MeV),
corresponding to a very small initial size $L\ll L_c$ ($L_c=0.5 - 1\;fm$), is
assumed to produce
a  $q \bar q$ pair which initiates a cascade of sequential gluon emissions.
The early stage is characterized by emission of ``hot'' gluons
far off mass shell in the perturbative vacuum. Subsequent gluon
branchings yield ``cooler'' gluons with successively
smaller virtualities, until their mutual separation approaches $L\approx
L_\chi$.
As is evident from the previous Fig. 2, this point characterizes the beginning
of the transition,
because the partons can now tunnel through the developing
potential barrier and form color-singlet composite states, which represent
the particle excitations of the scalar long-range $\chi$ field.
These ``pre-hadronic'' excitations must then convert
into physical hadronic states $-$ either  excited gluonic states, or, via
coupling to
the $U$ field, quark-antiquark meson excitations $-$
and subsequently decay into low-mass hadrons.
\bigskip

\noindent {\bf 4.1 General concept}
\medskip

As stressed in Sec. 2, the phenomenological
color-singlet function $\chi$ represents
the effect of the long-range order of the non-perturbative vacuum,
so that the formation of a color-neutral parton cluster - or bubble -
can be interpreted as a domain structure immersed in the medium of
the non-perturbative vacuum.
In quantum field theory such stable field configurations
arise as {\it classical} soliton solutions of the equations
of motion \cite{leebook,coleman85}.
On the other hand, it is well known that QCD exhibits the so-called
`preconfinement' property \cite{preconf} already on the perturbative level,
which is the tendency of the gluons and quarks produced in parton
cascades to arrange themselves in color-singlet clusters with limited
extension in phase space. It is therefore natural
to suppose that these clusters, or bubbles, are the basic `pre-hadronic' units
out of
which hadrons arise non-perturbatively.

Thus, the kinetic evolution of the system develops in three stages: parton
multiplication,
parton-cluster conversion, and cluster decay into hadrons.
It is clear that in this approach the conversion process is a local,
microscopic mechansim,
that proceeds earlier or later at different points in space,
depending on the local density of partons and their
nearest-neighbour separation $L$, as defined by eq. (\ref{Ldef}).
Thus, in order to trace the full dynamics,
it is necessary to
follow the evolution of the particle distributions in real time using the
kinetic framework of Sec. 3.
In accord with the above picture, we will now proceed in several steps,
starting from the master equations (\ref{T}) and (\ref{R}):
(i) employ  a separation of a coherent (mean) field part and
a contribution from quantum excitations for the composite
fields $\chi$ and $U$,
(ii) fix a specific, ghost-free gauge for the gluon fields
that is most convenient for our purposes,
(iii) treat the
evolution of the high-momentum quarks and gluons perturbatively in the
presence of the coherent field $\chi$.
\medskip

\noindent
{\bf(i)}
According to our interpretation of oscillations about the
minimum of the potential at $\chi_0, U_0$ as physical
excitations of the coherent fields $\chi$ and $U$, we
separate in a standard way \cite{fetter}
the classical field configuration at the minimum of ${\cal V}_{\chi=\chi_0}$ in
Fig. 2, from the quantum fluctuations around this minimum.
We  represent
$ \chi=\overline{\chi}+\hat\chi$ and $U=\overline{U} + \hat U$,
where $\overline{\chi}$, $\overline{U}$ are c-number functions
(the mean field parts), and $\hat\chi$, $\hat U$ denote quantum operators
(describing the excitations).
The physics behind this separation is that the coupling of $\chi$ to
$\psi$, $\overline{\psi}$, $A^\mu$, as well as to $U$, will make the
composite fields
$\chi$ and $U$ dynamical variables, so that the fluctuations around
the mean fields $\overline{\chi}$ and $\overline{U}$
will propagate and form collective excitations. Therefore the system
is characterized (aside from the elementary fields $\psi$, $\overline{\psi}$,
$A^\mu$)
by the mean fields, as well as by the collective excitations with
their own energy spectrum and distribution.
With this prescription we can treat the local interaction of the
partonic fluctuations with the coherent field analogously to the familiar
problem
of quantum fields ($\psi,\overline{\psi},A^\mu$)
interacting with a classical ``external'' field ($\overline{\chi}$),
which converts the partons to color-singlet clusters or
bubbles corresponding to excitations in the coherent field ($\hat \chi$).
Specifically, in our approach
the bubbles represent non-topological soliton configurations
which are stable, classical solutions of the equations of motions,
as have been studied for instance by Friedberg and Lee \cite{leebook}
and Coleman \cite{coleman85}.
We do not include here the possible additional interactions between partons
and bubbles, or among bubbles themselves.
\medskip

\noindent
{\bf(ii)}
It is convenient to work in  a  physical (axial) gauge \cite{dok80,gaugebook}
for the gluon fields, generically given by choosing the gauge
function $\xi_a(A)$ in (\ref{LQCD}) as
\begin{equation}
\xi_a(A)\;=\;
-\;\frac{1}{2\alpha n^2}\;\partial_\lambda (n\cdot A_a)\partial^\lambda (n\cdot
A_a)
\;,
\label{gauge}
\end{equation}
where $\alpha$ is the gauge parameter, and $n^\mu$ is a constant vector with
$n^2 \ne 0$.
In particular, we will set $\alpha=1$ which is known as the planar gauge.
In contrast to covariant gauges where $\xi_a(A)=-1/(2\alpha) (\partial\cdot
A_a)^2$,
the class of gauges (\ref{gauge}) is well known to have a number of advantages.
It is ghost-free, i.e. the ghost field
contribution in (\ref{genf3}) decouples and drops out.
Also, the so-called Gribov ambiguity is not present in this gauge.
Another feature of (\ref{gauge}) is that the gluon propagator involves only the
two physical transverse polarizations, so that the equations (\ref{T}) and
(\ref{R}) simplify considerably \cite{ms39}.
Furthermore, it allows for a rigorous resummation of the perturbative series
at high energies in terms of the leading logarithmic contributions and
consequently leads to a
simple probabilistic description of the perturbative parton evolution within
the (Modified) Leading Log approximation (MLLA) \cite{MLLA1,MLLA2} in QCD.
\medskip

\noindent
{\bf(iii)}
We will evaluate iteratively the 2-point Green functions ${\cal S}$ and ${\cal
D}$
of quarks and gluons, respectively, in the one-loop approximation
in the framework of ``jet calculus'' \cite{konishi79}, using the
MLLA techniques of coherent parton evolution including soft-gluon interference
\cite{MLLA1}.
The associated quark and gluon self energies $\Sigma$ and $\Pi$ include both
the
one-loop quark-gluon self interaction through real and virtual emission and
absorption,
and the effective interaction with the
confining background field $\overline{\chi}$ \cite{ms36}.
Similarly, the self-energy $\Xi$ of the $\chi$ field embodies
the self interaction and the coupling to the $U$ field,
as contained in the effective potential (\ref{V}),
as well as contributions from quark and
gluon recombination to $\chi$ excitations.
Correspondingly,
the function $\tilde{\Xi}$ of the $U$ field incorporates
its self interaction and the interaction with the $\chi$ field.
\bigskip

\noindent {\bf 4.2 The kinetic equations for real-time evolution in phase
space}
\medskip

As a consequence of the prescriptions  (i)-(iii),
and of exploiting  in the present  $e^+e^-\rightarrow hadrons$ case
the special property of translation invariance  of
the parton  evolution in the perturbative vacuum,
one finds after a lengthy calculation \cite{ms39} that the
transport equations (\ref{T}) and the
constraint equations (\ref{R}) can be combined in a single
set of coupled integro-differential equations for the phase-space densities
$F_\alpha (r,p)$ defined by (\ref{WW}) and (\ref{WWW}).
Introducing the usual light-cone variables
\begin{equation}
p^\mu \;=\;(p^+,p^-,\vec p_\perp) \;,
\;\;\;\;\;\;\;\;\;\;\;\;
p^\pm\;=\; p_0\,\pm\,p_z
\;,\;\;\;\;\;
\vec p_\perp \;=\; (p_x,p_y)
\end{equation}
and
\begin{equation}
x \;=\;
\frac{p^+}{Q}
\;\;,\;\;\;
p_\perp \;=\; \sqrt{p_x^2+p_y^2}
\;\;,\;\;\;
p^2 \;=\; p_\mu p^\mu
\;.
\end{equation}
where $Q$ is the hard scale of the initial $q\bar q$ pair created by
the photon, and $r\equiv r^\mu =(t,\vec r)$,
we write
\begin{equation}
F_\alpha \equiv F_\alpha(r,p) \;=\; F_\alpha(t,\vec r;x,p_\perp^2,p^2)
\;.
\end{equation}
The kinetic equations which one obtains from the transport equations (\ref{T})
by implementing
the constraints (\ref{R}) can now be summarized compactly as follows (see Fig.
6):
\begin{eqnarray}
\hat {\cal K} \; F_q
&=&
+\;\hat A_q^{qg}  \,F_q
\; + \;
\hat A_g^{q\bar q} \,F_g
\;-\;
\hat B_{qg}^{q\chi}\,F_q F_g
\;-\;
\hat B_{q\bar q}^{\chi\chi}\,F_q F_{\bar q}
\label{eq}
\\
\hat {\cal K} \; F_g
&=&
+\;\hat A_g^{gg}  \,F_g
\; - \;
\hat A_g^{q\bar q} \,F_g
\; + \;
\sum_f\,
\hat A_q^{g q} \,F_{q+\bar q}
\;-\;
\hat B_{gg}^{\chi\chi}\,F_g F_g
\;-\;
\sum_f\,
\hat B_{gq}^{\chi q}\,F_g F_{q+\bar q}
\label{eg}
\\
\hat {\cal K} \; F_\chi
&=&
+\;
\hat C_{gg}^{\chi\chi}\,F_g F_g
+
\sum_f\,
\hat C_{gq}^{\chi q}\,F_g F_{q+\bar q}
+
\sum_f\,
\hat C_{q\bar q}^{\chi\chi}\,F_q F_{\bar q}
-
\hat D_\chi^{U}\,F_\chi
-
\hat E_\chi^{h}\,F_\chi
\label{ec}
\\
\hat {\cal K} \;  F_U
&=&
\;+\;
\hat{D'}_{\chi}^{U}\,F_\chi
\;-\;
\hat{E'}_{U}^{h}\,F_U
\;\;,
\label{eu}
\end{eqnarray}
where we have abbreviated $F_{q+\bar q}\equiv F_q+F_{\bar q}$,
which in the present case is equal to $2 F_q$, and the
$\sum_f$ means summing over quark flavors $f=u,d,s,\ldots$.
%Pauli blocking and Bose enhancement effects are neglected.
The left-hand sides of these equations describe
the propagation of the particles in the presence of
the mean field, whereas the terms on the right-hand sides
represent the effects of particle creation, annihilation, and
recombination.
The operator $\hat {\cal K}$ on the left-hand sides is given by
\begin{equation}
\hat {\cal K} \;F_\alpha \;\,\equiv\;\,
\left[\frac{}{}
p_\mu \,\partial_r^\mu
\;+\; (\overline{M}_\alpha \,\partial_r^\mu \overline{M}_\alpha)
\,\partial_\mu^p
\right] \; F_\alpha
\;,
\end{equation}
with the first term describing the free propagation and the second
term reflecting the effect of the mean field (Fig. 6a). The functions
$\overline{M}_\alpha$ are the mean-field parts of the
self energies $M_\alpha$ defined in (\ref{WWWW}).
On the right-hand side the quantities
$\hat I_{a_1,a_2,..}^{b_1,b_2, ..}$ (where $\hat I = \hat A, \hat B, ..$)
represent integral operators
that incorporate the effects of the self energies in terms of
the relevant amplitudes for the various
processes $a_1,a_2,..\rightarrow b_1,b_2, ..$, and
that act on the phase-space densities to their right (Fig. 6b).
These coupled equations reflect a probabilistic
interpretation of the evolution in terms of successive
branching and recombination processes,
in which the changes of the particle distributions on the left-hand sides
are governed by the balance of gain (+) and loss ($-$) terms on the right-hand
sides.
The different terms on the right-hand sides of eqs. (\ref{eq})-(\ref{eu}), the
contributions
to the gain and loss of particles, fall into three categories:

\noindent
{\bf (a)}
Parton multiplication through emission processes
$q\rightarrow qg$, $g\rightarrow gg$ and $g\rightarrow q\bar q$;

\noindent
{\bf (b)}
Parton cluster formation through recombinations
$q\bar q\rightarrow \chi\chi$, $qg\rightarrow q\chi$, $gg\rightarrow \chi\chi$;

\noindent
{\bf (c)}
Hadronic cluster decay either through direct conversion
of the formed scalar $\chi$ excitations
into hadrons $h$ through $\chi\rightarrow h$, or via coupling to the
pseudoscalar
states $\chi\rightarrow U$,  and the subsequent decay into hadrons,
$U\rightarrow h$.
\smallskip

\noindent
In the following subsections we explain these contributions in detail.
\medskip

\noindent {\bf 4.3 Parton multiplication}
\medskip

The integral operators $\hat A$ in the
quark and gluon evolution equations (\ref{eq}) and (\ref{eg})
represent the changes of the parton distributions in phase space
due to the perturbative cascade evolution.
Explicitly,
\begin{eqnarray}
\hat A_q^{qg}  \,F_q
&=&
\lambda_\chi \,
\int_0^1 dz \,\left[\frac{1}{z} \, F_q\left(r;\frac{x}{z},z p_\perp^2,z
p^2\right) \;-\;
 \,F_q(r; x,p_\perp^2,p^2)\right]\; \gamma_{q\rightarrow
qg}\left(z,\epsilon\right)
\;a_q(z,p^2)
\nonumber \\
\hat A_g^{q\bar q}  \,F_g
&= &
\lambda_\chi \,
\int_0^1 \frac{dz}{z}  \, F_g\left(r;\frac{x}{z},z p_\perp^2, z p^2\right)
 \, \gamma_{g\rightarrow q\bar q}\left(z,\epsilon\right)
\;a_g(z,p^2)
\nonumber \\
\hat A_g^{gg}  \,F_g
&= &
\lambda_\chi \,
\int_0^1 dz \,\left[\frac{1}{z} \, F_g\left(r;\frac{x}{z},z p_\perp^2,z
p^2\right) \;-\;
\frac{1}{2} \; \,F_g(r; x,p_\perp^2,p^2)\right]\; \gamma_{g\rightarrow
gg}\left(z,\epsilon\right)
\;a_g(z,p^2)
\nonumber \\
\hat A_g^{q\bar q}  \,F_g
&= &
\lambda_\chi \,
n_f(p^2)\; F_g(r;x,p_\perp^2, p^2) \;
\int_0^1 dz \, \gamma_{g\rightarrow q\bar q}\left(z,\epsilon\right)
\;a_g(z,p^2)
\nonumber \\
\hat A_q^{gq}  \,F_{q+\bar q}
&= &
\lambda_\chi \,
\int_0^1 \frac{dz}{z} \,  F_{q+\bar q}\left(r;\frac{x}{z},z p_\perp^2, z
p^2\right)
\; \gamma_{q\rightarrow gq}\left(z,\epsilon\right)
\;a_q(z,p^2)
\;.
\label{Ta}
\end{eqnarray}
Here
$\lambda_\chi \equiv \lambda_\chi(\chi(r))= 1 - (\chi/\chi_0)^4 +
O[(\chi/\chi_0)^6]$,
and the function $a(z,p^2)$ is given by
\begin{equation}
a_{q,g}(z,p^2) \;:=\;
\frac{1}{2 \pi}\; T_{q,g}(p^2)\;
\alpha_s\left(\frac{}{}(1-z)p^2\right)
\;,
\end{equation}
with a ``life-time'' factor $T_{q,g}(p^2)$ that expresses the probability
for a parton of virtuality $p^2$ to decay (branch) within
a time interval $t$ in the laboratory frame,
\begin{equation}
T_{q,g}(p^2)\;=\; 1\;-\; \exp\left[ - \frac{t}{\tau_{q,g}(p^2)}\right]
\;,
\label{TLF1}
\end{equation}
where $\tau(p^2) \propto E/p^2$ (explicit expressions can be found
in Ref. \cite{ms1}). Furthermore,
$\alpha_s$ is the one-loop QCD coupling
\begin{equation}
\alpha_s (k^2)\;=\;
\frac{12\pi}{\left(\frac{}{} 33 -2 n_f(k^2)\right)\,
\ln\left[(k^2+k_0^2)\,L_c^2 \right]}
\;,
\label{alpha}
\end{equation}
and $n_f(k^2)$ is the effective number of quark flavors at
$k^2$,
\begin{equation}
n_f(k^2)\;:=\;\sum_f^{N_f}\,\sqrt{1-\frac{4 m_f^2}{k^2}}
\;\theta\left(1 - \frac{4 m_f^2}{k^2}\right)
\;.
\end{equation}
In (\ref{alpha}) we have assumed the correspondence
$L_c \simeq \Lambda^{-1}_{QCD}$ to the intrinsic perturbative QCD scale,
and $k_0$ is a parameter that prevents a divergence
when $k^2 \rightarrow L_c^{-2}$, and defines a maximum value
$\alpha_s(0)$. We will determine $k_0$ in Sec. 5 from the total parton
multiplicity.
The functions $\gamma(z,\epsilon)$ are are analogous
to the standard branching kernels in the MLLA \cite{dok80}.
Note that the 4-gluon vertex vertex does not contribute in the
MLLA in the gauge (\ref{gauge}), because it is kinematically suppressed.
As a consequence, the effect of the couplings $\kappa_L (\chi)$ and
$\mu(\chi)$, eqs. (\ref{kappa}) and (\ref{Mchi}),  on the
parton evolution reduces to  2-parton recombinations into
color-singlet clusters $-$ the terms proportional to $\hat B, \hat C$ which
will be given below.

The branching kernels $\gamma_{a\rightarrow bc}(z)$ are
the familiar energy distributions for the branching $a\rightarrow bc$
with $z = x_b/x_a$ and $1-z = x_c/x_a$ the energy fractions of
of daughter partons:
\begin{eqnarray}
\gamma_{q \rightarrow q g} (z,\epsilon) &=&
C_F\; \left( \frac{1 + z^2}{1 - z +\epsilon} \right)
\nonumber
\\
\gamma_{q \rightarrow g q} (z,\epsilon) &=&
C_F\; \left( \frac{1 + (1-z)^2}{z + \epsilon} \right)
\nonumber
\\
\gamma_{g \rightarrow g g} (z,\epsilon) &=&
2\,C_A\;\left( \frac{z}{1-z+\epsilon} + \frac{1-z}{z + \epsilon} + z ( 1 - z )
\right)
\nonumber
\\
\gamma_{g \rightarrow q \bar q} (z,\epsilon) &=&
\frac{1}{2} \, \left( z^2 + (1 - z)^2 \right)
\;\;,
\label{gamma}
\end{eqnarray}
where $C_F = (N_c^2-1)/(2 N_c) = 4/3$, $C_A = N_c =3$.
In the denominator of $\gamma_{q\rightarrow qg}$ and $\gamma_{g\rightarrow
gg}$,
there appears the function
\begin{equation}
\epsilon \;=\; \frac{p^{'\,2} n^2}{4 (p\cdot n)^2} \;\propto \frac{
p_\perp^2}{p_z^2}
\;,
\end{equation}
where $p$ $(p')$ is the momentum of the mother (daughter) parton
and $p_\perp$ the relative transverse momentum of the daughter partons
with respect to the mother. It
arises here as a consequence of the constraint equations (\ref{R})
which determine spatial uncertainty associated with
the off-shellness of  the partons.
It effectively cuts off small-angle gluon emission
by modifying the free gluon propagator $\propto z_g^{-1}$ to the form
$(z_g+\epsilon)^{-1}$
(where $z_g = z$ or $z_g=1-z$) when $p_\perp/p_z = O(1)$, that is, in branching
processes with
large space-time uncertainty. This
ensures that the two daughter partons can be resolved as
individual quanta only if they are separated sufficiently by
$\Delta r_\perp \propto 1/p_\perp$ in position space,
in accord with the uncertainty principle.
Note that $\epsilon$ can be neglected in the terms
$\propto (z_g+\epsilon)^{-1}$ in (\ref{gamma})
for energetic gluon emission ($z_g\rightarrow 1$), but is essential in the
soft regime $(z_g\rightarrow 0$).
The effect of $\epsilon$ has been shown \cite{dok80,amati80} to result in a
natural
regularization of the infra-red-divergent behaviour of
the branching kernels (\ref{gamma}), due to destructive gluon
interference which becomes complete in the limit $z_g\rightarrow 0$.
\bigskip

\noindent {\bf 4.4 Parton cluster (bubble) formation}
\medskip

The operators $\hat B, \hat C$ in eqs. (\ref{eq})-(\ref{ec})
represent the changes of the phase-space densities due to
recombinations of  two partons at $r$ and $r'$ to  color-neutral
clusters, or bubbles that arise as non-trivial structures
in the vacuum because of the confinement mechanism.
Their formation is determined, in analogy to the finite-temperature
QCD phase transition \cite{CEO}, by the probability for tunnelling
through the potential barrier of ${\cal V}$ between $\chi=0$ and
$\chi=\chi_c$ in Fig. 2,
which separates perturbative and non-perturbative vacua.
The associated rate of bubble formation around $L=L_c$ is generically given by
an
exponential probability distribution \cite{langer69,ellis93},
\begin{equation}
\pi(L)\;=\; \pi_0(L)\;\left(\frac{}{}1\;-\; \exp[-\Delta F\;L]\right)
\;,
\label{piV}
\end{equation}
where $\pi_0= const. \ln(1-u) \theta(1-2u)+\theta(2u -1)$,
with $u=\Delta F \,L$,
modifies the small-$L$ behaviour for which the
exponential form is not appropriate.
Here $\Delta F$ is the change in the free energy
of the system that is associated with the conversion from partons to clusters.
In our case (for baryon-free matter in general),
\begin{equation}
\Delta F\; \;=\;
E_{vol}\;+\;E_{surf}
\;=\;
\frac{4\pi}{3}\;R^3(L)\;\Delta P(L)\;+\;4\pi\;R^2(L)\;\sigma(L)
\;,
\label{DF}
\end{equation}
where $R(L)$ is the radius of the bubble depending on the parton separation.
The first term is the volume energy determined by
the difference of pressure in the perturbative  and the non-perturbative
vacuum,
\begin{equation}
\Delta P(L)\;=  P_{qg}(L)\;- P_\chi(L)
\;.
\end{equation}
The second term in (\ref{DF}) is the surface energy with
the surface tension estimated to be
\begin{equation}
\sigma(L)\;=\;\int_{0}^{\chi_c}d\chi \;\sqrt{2\,{\cal V}(L)}
\;\approx\;2\;\int_{\chi_{max}}^{\chi_c}d\chi \;\sqrt{2\,{\cal V}(L)}
\label{sigma}
\end{equation}
where $\chi_c$ corresponds to the local minimum of ${\cal V}$
at $L_c$ and $\chi_{max}$ is the point of the local maximum
of ${\cal V}$ that separates unconfined and confined domains, as defined in
Fig. 2.

A parton bubble is stable if $\partial \Delta F /\partial R= 0$,
which leads to the condition for the stationary bubble radius,
\begin{equation}
R_c\;\equiv\; R(L_c) \;=\; \left.\frac{2 \,\sigma}{\Delta P}\right|_{L=L_c}
\end{equation}
When inserted in (\ref{DF}), this gives for (\ref{piV})
\begin{equation}
\pi(L)\;=\; \pi_0(L) \;
\left(\frac{}{}1\;-\; \exp\left[-\frac{4\pi}{3}\,R_c^2 \sigma_c
\,L\right]\right)
\label{piV2}
\end{equation}
with $\sigma_c \equiv \sigma(L_c)$.
In accord with our definition (\ref{Ldef}),
we interpret the space-time scale $L$ as
the characteristic inter-parton separation, that is, we define it
in terms of the  distance measure $\Delta_{i j}$
between two partons, labeled with indices $a$ and $b$,
\begin{equation}
\Delta_{ab} \;=\; \sqrt{ r_{ab}^\mu \; r_{ab, \mu} }
\;\;,\;\;\;\;\;
r_{ab}\; = \;r_a\, -\, r_b
\;,
\label{Dab}
\end{equation}
and identify $L$ with
the the minimum distance $L_{ab}$ for a certain parton $a$
to its next neighbour $b$:
\begin{equation}
L(r)\;=\; L_{ab} \;\equiv \;
\mbox{min}_{\;b} (\Delta_{a 1}, \ldots , \Delta_{a b}, \ldots , \Delta_{a n})
\;.
\label{L}
\end{equation}
Other measures are possible, e.g.
$\Delta_{ab}= (1/x_{ab}^2+1/y_{ab}^2+1/z_{ab}^2+1/(\Delta t)^2)^{-1/2}$,
but we find that the particular choice of $\Delta_{ab}$ is not crucial as long
it provides a reasonable distance measure.
We choose (\ref{Dab}), because it  has the advantage that it is manifestly
Lorentz-invariant and
has a simple interpretation as the two partons' spatial separation
in their center-of-mass frame.

We assume that the dominant contribution to bubble formation arises from
2-parton fusion and ignore recombinations of 3 or more partons.
This is reasonable, unless the local parton density becomes so large that
also the latter processes have a significant probability to occur.
We take the same probability distribution $\pi(L)$ for
the various types of configurations, since it depends only on the `color- and
flavor-blind' variable $L$, i.e. we set
\begin{equation}
\pi(r,r')\;=\;\pi(L)\;\equiv\; \pi_{gg\rightarrow \chi\chi}(L)\;=\;
\pi_{q\bar q\rightarrow \chi\chi}(L)\;=\;
\pi_{gq\rightarrow \chi q}(L)
\;,
\end{equation}
where $L$ is given in terms of $r$ and $r'$ by (\ref{L}).
The various $\hat B$ terms in (\ref{eq}) and (\ref{eg})
can then be expressed generically as:
\begin{equation}
\hat B_{ab}^{\chi c}  \,F_a F_b
\;=\;
F_a(r;x,p_\perp^2,p^2)
\; \int d^3 r' \,\pi(r,r') \int d^4p'\, F_b(r';x',p_\perp^{'\,2}, p^{'\,2})
\label{Tb}
\end{equation}
where $a,b = q,\bar q, g$ and $c=\chi, q$, and
$d^4 p'= dp^{'\,2} dp^{'\,2}_\perp dx'/x'$.
Similarly the $\hat C$ terms in (\ref{ec}) are given by:
\begin{eqnarray}
\hat C_{ab}^{\chi c}  \,F_a F_b
&=&
\int d^3 r' \int d^3 r''\,\pi(r',r'')
\,\delta^3\left(\vec r-\frac{\vec r^{\,'}+\vec r^{\,''}}{2}\right)\,
\nonumber \\
& &
\; \times \;
\int d^4p'd^4p''\,
F_a(r';x',p_\perp^{'\,2},p^{'\,2})
F_b(r'';x'',p_\perp^{''\,2}, p^{''\,2})
\,\delta^4\left(p-\frac{p'+p''}{2}\right)
\label{Tc}
\;.
\end{eqnarray}
\medskip

\noindent {\bf 4.5 Hadronic cluster decay}
\medskip

The ensemble of clusters determined by the coupled equations
(\ref{eq})-(\ref{ec})
yields a continous mass spectrum of color-singlet excitations
with different flavor contents corresponding to the types of recombined
partons.
These states must then decay into physical hadronic states with a discrete mass
spectrum.
The invariant mass distribution of the formed clusters  may be interpreted
as a `smeared out' version of the spectrum of primordial resonances
formed in the early stages of the confinement mechanism \cite{MLLA1}.
It therefore seems reasonable to treat the fragmentation of these central
clusters
as a kind of averaged resonance decay which, as implied by our locality
assumption, must be determined entirely by their invariant masses, flavors
and total angular momenta.
Each cluster in the resonance spectrum may either represent a single
hadron resonance that converts directly into a physical hadron
with a definite mass, or else fragments through a two-body decay
into a pair of final-state hadrons. From the particle spectra obtained
in $e^+ e^-$-annihilation experiments it appears that
quasi-two-body final states are universally dominant, so that the
latter possibility seems favored if kinematically allowed.

We adopt the cluster fragmentation scheme of Refs. \cite{ms3,herwig},
however with some modification concerning heavy clusters. We assume that
each cluster $C=\chi, U$ can decay by either one of the following mechanisms:
\smallskip

\noindent {\bf (i)}
If a cluster $C$ is too light to
decay into two hadrons, it is taken to represent
the lightest single hadron (meson) $h$, corresponding to its
partonic constituents, $C  \rightarrow  h$,
with its mass shifted to the appropriate value by adjusting its energy
through exchange with a neighbouring cluster.
\smallskip

\noindent {\bf (ii)}
If, however, a cluster is massive enough to decay,
it decays isotropically in its rest frame into
a pair of hadrons (mesons or baryons),
$C \rightarrow  h_1 + h_2$
according to the decay probability specified below.
\smallskip

\noindent
Occasionally it occurs that
a cluster comes out very heavy, in which case isotropic 2-body decay
is not a reasonable mechanism any more. In this case we impose the constraint
that, if a cluster is heavier than a critical threshold $M_{crit}= 4$ GeV,
then it is rejected and the two recombining partons of that potential cluster
propagate on as individual quanta, and continue to participate in the
parton cascade process, either until they have decreased their virtuality
sufficiently,
or until they recombine with a lower-mass partner.

To implement this scheme, we
observe from eqs. (\ref{ec}) and (\ref{eu}) that  the cluster-hadron
transformation
can proceed through the scalar channel $\chi\rightarrow h$, or
via the pseudoscalar channel
$\chi\rightarrow U\rightarrow h_1h_2$, depending on the corresponding density
of
states with masses below the decaying cluster.
We assume a Hagedorn \cite{hagedorn} density of states
\begin{equation}
\rho_h(m)\;=\; c\;m^{-a}\;\exp\left(-\frac{m}{T_0}\right)
\;,
\label{mspec}
\end{equation}
where $c$, $a$ are constants and $T_0$ is the
Hagedorn temperature with the typical values
$c=8 m_\pi^2$, $a=3$ and $T_0 = m_\pi$.
The  decay probability of a cluster with mass $m_C=\sqrt{p^2}$
to decay into a hadron state of mass $m'_h=\sqrt{p^{'\,2}}$ is then given by
\begin{equation}
\Gamma_{C\rightarrow h}(p,p')\;=\;\frac{1}{N_C(p^2)}
\;T_C(p^2) \;\int_{m'_h=\sqrt{p^{'\,2}}}^{m_C=\sqrt{p^2}} dm\;\rho_h(m)
\;.
\label{Gamma1}
\end{equation}
Here
\begin{equation}
N_C(p^2)\;=\;\int_{m_\pi}^{\sqrt{p^2}} dm\;\rho_h(m)
\;,
\end{equation}
and in analogy to (\ref{TLF1}), $T_C(p^2)$ is a
``life-time'' factor giving the probability
that a cluster of mass $m_C^2=p^2$ decays  within
a time interval $t$ in the laboratory frame,
\begin{equation}
T_C(p^2)\;=\; 1\;-\; \exp\left[ - \frac{t}{\tau_C(p^2)}\right]
\;,
\label{TLF2}
\end{equation}
where in this case we simply take $\tau_C(p^2)=E_C/p^2 = \gamma/m_C$
from the uncertainty principle.
In order to find the value for the decay probability
(\ref{Gamma1}) for a given cluster of mass $m_C$, we sum over
the possible decays for this cluster according to the particle data tables.
The probability for a specific 2-body decay mode is taken to be a product of
a flavor, a spin and a kinematical factor \cite{ms3},
\begin{equation}
\Gamma_{C\rightarrow h_1 h_2} ( m_C; m_1, m_2)
\;:=\;P_m(m_C, m_1+m_2) \; P_s ( j_1, j_2) \; P_k (m_C, m_1, m_2)
\;,
\end{equation}
where
$j_{1,2}$ ($m_{1,2}$) are the angular momenta (masses) of
the two hadrons $h_{1,2}$.
The factor
\begin{equation}
P_m ( m_C, m_1+m_2) \;=\; \left( 1 + \frac{m_1^2+m_2^2}{m_c^2} \right) \;
\sqrt{ 1 - \frac{(m_1+m_2)^2}{M_c^2}}\;\theta(m_C - m_1 - m_2)
\end{equation}
is the two-body phase-space suppression function for the decay.
The spin factor
\begin{equation}
P_s ( j_1, j_2 ) \;=\; \left( 2 j_1 + 1 \right) \, \left( 2 j_2 + 1 \right)
\end{equation}
takes into account the spin degeneracy with the allowed spins $j_1$ and $j_2$
of the two hadrons.
The kinematic factor
\begin{equation}
P_k ( m_C, m_1, m_2) \;=\; \frac{\sqrt{\lambda(m_C^2, m_1^2, m_2^2)}}{m_C^2}
\end{equation}
is the two-body phase-space factor, where
$
\lambda(a, b, c) = a^2 + b^2 + c^2 - 2 \left( a b + a c + b c \right)
$.

Thus, with the decay probability
$\Gamma_{C\rightarrow h}$ of (\ref{Gamma1}) evaluated in this fashion,
the terms involving the $\hat D$ and $\hat E$ operators in the kinetic
equations (\ref{ec}) and (\ref{eu})
can be expressed as
\begin{eqnarray}
\hat D_\chi^U  \,F_\chi
&=&
F_\chi(r;x,p_\perp^2,p^2)\;\int dp^{'\,2}\,\Gamma_{\chi\rightarrow
U}(p^2,p^{'\,2})
\nonumber \\
\hat E_\chi^{h}  \,F_\chi
&=&
F_\chi(r;x,p_\perp^2,p^2)\;\int dp^{'\,2}\,\Gamma_{\chi\rightarrow
h}(p^2,p^{'\,2})
\nonumber \\
\hat{D'}_\chi^U  \,F_\chi
&=&
\int d p^{'\,2} dp_\perp^{'\,2} \frac{dx'}{x'}
F_\chi(r;x,p_\perp^{'\,2},p^{'\, 2})\;\Gamma_{\chi\rightarrow U}(p^{'\,2},p^2)
\nonumber \\
\hat{E'}_U^{h}  \,F_U
&=&
F_U(r;x,p_\perp^2,p^2)\;\int dp^{'\,2}\,\Gamma_{U\rightarrow h}(p^2,p^{'\,2})
\label{Td}
\end{eqnarray}
\medskip

\noindent {\bf 4.6 Method of solution by Monte Carlo simulation}
\medskip

We can now solve the set of evolution equations (\ref{eq})-(\ref{eu}) by means
of a real-time simulation in full phase space using the computational methods
of Ref. \cite{ms3}.
One starts from an initial phase-space density of partons, which in the case of
a
jet-initiating $q\bar q$ pair with invariant mass $Q$ is
\begin{equation}
F_{q+\bar q}(t=0,\vec r; x, p_\perp^2,p^2)\;=\;
\delta^3(\vec r)\,\delta(\vec r^2 -
Q^2)\,\delta(x-1)\,\delta(p_\perp^2)\,\delta(p^2-Q^2)
\;,
\end{equation}
where
we choose the $q\bar q$ center-of-mass frame as our reference frame, and
we sum over all quark flavors $f$ weighted with a factor
$w_f=e_f^2/n_f(Q^2)\sqrt{1-4 m_f^2/Q^2}\,\theta\left(Q^2 - 4 m_f^2\right)$,
that accounts for the electromagnetic charge and mass threshold of the
initial $q\bar q$ pair produced by the photon (or $Z^0$).

The parton shower development is then followed in a cascade simulation
(for details see \cite{msrep,ms3}) in the center-of-mass frame:
The system of particles is evolved in discrete time steps, here taken
as $\Delta t = 0.01$ $fm$, in coarse-grained 7-dimensional phase-space with
cells $\Delta \Omega = \Delta^3 r \Delta^3 p \Delta M^2$.
The partons propagate along classical trajectories until they interact,
i.e., decay (branching) or recombine (cluster formation). Similarly, the
formed clusters travel along straight lines until they decay into hadrons.
The corresponding probabilities and time scales of interactions are
sampled stochastically from the relevant probability distributions according
to Secs. 4.3-4.5.
At any time $t>0$ we can extract the phase-space densities (\ref{WWW}),
$F_\alpha (t,\vec r,\vec p, p^2) = d N_\alpha(t)/d^3r d^4p$
of the particle species $\alpha=q,g,\chi,U$, and with these
phase-space profiles we can then calculate, using
the formulae (\ref{Tmunu}) and (\ref{pr}),
the associated local pressure $P(r)$, particle density $n(r)$
and energy density $\varepsilon(r)$
for each species individually, these being
the quantities that characterize the macroscopic state
of the system at $t$ and $\vec r$.

With this concept, we can trace the space-time evolution of the
parton-hadron conversion process self-consistently:
at each time step, any ``hot'' off-shell parton is allowed
to decay into ``cooler'' partons, with a probability determined
by its virtuality and life time. Also in each step, every parton and its
nearest
spatial neighbor are considered as defining a fictious space-time bubble
with invariant radius $L$, as defined by (\ref{L}).
By comparing the local pressure of partons $P_{qg}(t,\vec r,L)$ with the
pressure $P_{\chi}(t,\vec r,L)$  that such a pre-hadronic bubble would
create instead, we obtain the associated value of the
conversion probability $\pi(L)$ which determines the cluster (bubble)
formation rate explained in Sec. 4.5.
If the partons do convert into a cluster, they disappear from
a phase-space cell, and instead the composite cluster appears at the same
space-time point,
from which it propagates on.
Otherwise the partons continue in their shower development.
The final decay of each formed cluster into hadrons is simulated analogously,
except that it does not require the comparison of pressures, but
is determined by kinematics and the available phase space.
This cascade evolution is followed until all partons
have converted, and all clusters have decayed into final-state hadrons.

As an illustrative example, we show in Fig. 7 the time evolution of the
particle density
profiles of partons (Fig. 7a) and clusters (Fig. 7b) for a jet system with
invariant mass
$Q=100$ GeV. It is evident how the system evolves in position space with
respect to
the center-of-mass of the two inital partons as a polar wave front
(the pictures are symmetric in $r_\perp =\sqrt{r_x^2+r_y^2}$), with the partons
gradually converting
to clusters.
It is interesting that this  local excitation of the vacuum due to the
injection of
the highly virtual dijet system, and the subsequent evolution, resemble very
much
the situation of a stone plunged into water, with a well-shaped ``shock front''
expanding isotropically in the center-of-mass frame.
\bigskip

\bigskip
\noindent {\bf 5. PHENOMENOLOGY}
\bigskip

In this Section we study the observable implications of our approach to
parton-hadron conversion, and investigate its
consistency with standard particle physics phenomenology.
\bigskip

\noindent {\bf 5.1 Determination of the potential {\boldmath $V(\chi,U)$}}
\medskip

We first need to specify the parameters of our approach.  Recall that
this phenomenological input is contained in the effective long-range
potential $V(\chi,U)$, eq. (\ref{V}), which combines with the
dynamical contribution $\delta V(L,\chi)$ to the $L$-dependent
potential ${\cal V}(L)$, eq. (\ref{calV}).
As $L$ varies, ${\cal V}$ changes its shape, which affects the dynamical
evolution of the system, and the latter in turn determines the further
change of ${\cal V}$.
Hence, the details of the dynamics are governed by the choice of
parameters in $V$ and thus ${\cal V}$.
The crucial parameters are the bag constant $B$ which defines the
vacuum pressure $V(0)$ in the short-distance limit $L\rightarrow 0$,
and $\chi_0$, the value of the condensate of $\chi$ in the long-distance
regime.
As indicated in Fig. 2, as $L$ increases, the changing form of ${\cal V}(L)$ is
characterized by three distinct length scales:
$L=L_\chi$, the point when partons begin to
convert, $L= L_c$, when the pressures of partons and pre-hadronic clusters
equal each other, and $L=L_0$, when the transition is completed.

We will fix $B$ and $\chi_0$, which have  well-defined physical
interpretations,
and then determine $L_\chi$, $L_c$ and $L_0$.
Although the values of $B$ and $\chi_0$ are not precisely known, there
is agreement of various phenomenological determinations about their ranges:
one expects \cite{CEO} $B^{1/4}=(150 - 250)$ MeV and
$\chi_0=(50 - 200)$ MeV. In the following we adopt two
representative parameter combinations:
\begin{eqnarray}
B^{1/4} \;=\;230\;\mbox{MeV}
& &
\;\;\;\;
\chi_0 \;=\;200\;\mbox{MeV}
\nonumber \\
B^{1/4} \;=\;180\;\mbox{MeV}
& &
\;\;\;\;
\chi_0 \;=\;100\;\mbox{MeV}
\;.
\label{paraval}
\end{eqnarray}
Then, with the potential ${\cal V}$ specified, we can determine
the values of $L_\chi$, $L_c$, $L_0$ from the Monte Carlo simulation
of the evolution of the system as the scale $L$ changes due to the particles'
diffusion
in phase space.
The most interesting quantity is $L_c$, the point which is characterized
by the equality of partonic and hadronic pressures. As explained in Sec. 4.6,
we can compute the corresponding pressures
$P_{qg}$ and $P_\chi$ from the phase-space densities of partons and
clusters, respectively.
In analogy to Ref. \cite{CEO}, we represent (on dimensional grounds)
\begin{eqnarray}
P_{qg}(r,L) &=& a_{qg}(r,L)\;L^{-4} \;\,-\;\, B
\nonumber \\
P_{\chi}(r,L) &=& a_\chi(r,L)\;L^{-4}\;\,-\;\, {\cal V}(L)
\;,
\end{eqnarray}
and, because ${\cal V}(\chi,L)\vert_{L=L_c}={\cal V}(\chi_c,L_c)=0$ (c.f. Fig.
2), we have
\begin{equation}
L_c\;=\; \left[\frac{a_{qg}(r,L_c)\;-\;a_\chi(r,L_c)}{B}\right]^{1/4}
\label{Bval}
\;.
\end{equation}
The dimensionless functions $a_{qg}$ and $a_\chi$ are obtained from the
numerical simulation and are shown in Fig. 8 as a function of time for
the above two choices of $B$ and $\chi_0$ in the cases of $q\bar q$ and $gg$
jet evolution with $Q=10$ GeV and $Q=100$ GeV.
Plotted are the kinetic pressures $P(t,L):= a(t,L) L^{-4}$ (where $t \simeq
\sqrt{L^2+z^2}$) along the
``shock front'' of the jet profile which is seen in the previous Fig. 7.
{}From Figs. 8a and 8b one observes that
(i) the pressure evolution obviously  depends on the type of
the two jet-initiating partons: it decreases more slowly for $q\bar q$ pairs of
different flavors than
for a $gg$ pair, because gluons have a larger emission rate and therefore the
two
leading gluons evaporate their initial energy faster;
(ii) the crossover point between the pressures  $P_{qg}$ and $P_\chi$
is rather insensitive to the choice of $L_c$; (iii) the crossover is shifted
away from $t=0$ with
increased jet energy $Q$;  (iv) at $L_c$ the partonic pressure
$P_{qg}$ still exceeds $P_\chi$, i.e.
$a_{qg}(L_c) > a_\chi(L_c)$, consistent with (\ref{Bval}).
{}From this analysis, we find using the determining condition (\ref{Bval}),
\begin{equation}
L_c\;=\;\left\{
\begin{array}{c}
0.6 \; fm \;\;\mbox{for}\; B^{1/4}\;=\;230\;MeV \\
0.8 \; fm \;\;\mbox{for}\; B^{1/4}\;=\;180\;MeV
\end{array} \right.
\;.
\end{equation}

Directly associated with the scale $L_c$ is the
parton-cluster conversion probability (\ref{piV2}), which
is determined by the width of the potential wall between the
two phases. It enters the kinetic equations via (\ref{Tb}) and
determines locally the time scale of the parton-to-cluster transition by
the magnitude of the surface tension $\sigma_c$ as given by (\ref{sigma}).
We find
\begin{equation}
\sigma_c^{1/3}\;=\;\left\{
\begin{array}{c}
40\;MeV \;\;\mbox{for}\; L_c\;=\;0.6\;fm \;\; (B^{1/4}=230\;MeV)\\
48\;MeV \;\;\mbox{for}\; L_c\;=\;0.8\;fm \;\; (B^{1/4}=180\;MeV)
\end{array} \right.
\label{sigval}
\;,
\end{equation}
which by virtue of (\ref{piV2}) fixes the cluster-formation probability
$\pi(L)$.
It is noteworthy that the above  small values of the surface tension
$\sigma_c$ are  in agreement with lattice QCD simulations \cite{ignatius94}.
and correspond to a weakly
first-order transition at finite temperature, which is consistent
with astrophysical constraints \cite{thomas93} on inhomogenities.
This finding implies a rather rapid conversion of partons into color-singlet
clusters (pre-hadrons), as is also evident from Fig. 8.
This means that parton-hadron conversion is not dependent on the details
of the interpolation functions $\kappa_L(\chi)$ and $\mu_L(\chi)$, as already
advertised in Sec. 2.3,
and has interesting consequences for the cluster size and mass distribution,
as we will discuss below.

The value of $L_\chi$,
below which size only the perturbative vacuum of the pure parton phase can
exist,
is given by the point of inflection of the effective potential ${\cal V}$, when
the
local minimum at $\langle \chi \rangle \ne 0$ ceases to exist (c.f. Fig. 2).
It turns out to be rather close to $L_c$,
\begin{equation}
L_\chi\;\approx\;\left\{
\begin{array}{c}
0.4-0.5\;fm \;\;\mbox{for}\; L_c\;=\;0.6\;fm \;\;(B^{1/4}=230\;MeV) \\
0.6-0.7\;fm \;\;\mbox{for}\; L_c\;=\;0.8\;fm \;\;(B^{1/4}=180\;MeV)
\end{array} \right.
\;,
\end{equation}
which is a consequence of the  small values (\ref{sigval})
for the surface tension $\sigma_c$,
and indicates that the transition occurs very abruptly. Finally,
we find that the scale $L_0$, when the parton-cluster conversion is complete
and no partons are left over, depends not only on $L_c$ but also
on the initial jet energy $Q$. It gives an estimate for the time scale
$\tau_0\propto L_0$ of the global conversion process, and
comes out rather long, namely for $L_c=0.6$ $fm$ we get
$\tau_0 \approx 8.5$ (21) $fm$ for $Q=10$ (100) GeV, whilst
for $L_c=0.8$ $fm$ the time scale is
$\tau_0 \approx 10$ (26) $fm$ for $Q=10$ (100) GeV.
\smallskip

Immediate consequences of the values for $B$ and $\chi_0$ in (\ref{paraval})
are
the ``critical temperature'' for the phase transition in finite-temperature
QCD,
\begin{equation}
T_c\;\equiv\;
\left(\;\frac{9 \,B}{4 \pi^2}\right)^{1/4}\;=\;
\;\left\{
\begin{array}{c}
160\;MeV \;\;\mbox{for}\; L_c\;=\;0.6\;fm \;\;(B^{1/4}=230\;MeV) \\
125\;MeV \;\;\mbox{for}\; L_c\;=\;0.8\;fm \;\;(B^{1/4}=180\;MeV)
\end{array} \right.
\label{Tcval}
\;,
\end{equation}
the characteristic mass scale of the lightest scalar
glueball, given by \cite{CEO}
\begin{equation}
m_\chi\;\equiv\;\sqrt{\left.\frac{\partial^2 V(\chi,0)}{\partial
\chi^2}\right|_{\chi=\chi_0}}
\;=\;4\;\frac{\sqrt{B}}{\chi_0}\;=\;
\;\left\{
\begin{array}{c}
1.05\;GeV \;\;\mbox{for}\; L_c\;=\;0.6\;fm \\
1.30\;GeV \;\;\mbox{for}\; L_c\;=\;0.8\;fm
\end{array} \right.
\label{chival}
\;,
\end{equation}
and, the estimate for the value of the gluon condensate (\ref{gcond}),
\begin{equation}
G_0\;=\;\frac{32}{9}\;B\;=\;
\;\left\{
\begin{array}{c}
1.25\;GeV\,fm^{-3} \;\;\mbox{for}\; L_c\;=\;0.6\;fm \\
0.50\;GeV\,fm^{-3} \;\;\mbox{for}\; L_c\;=\;0.8\;fm
\end{array} \right.
\;.
\end{equation}

\noindent
The parameter values obtained above are summarized in Table 1. Both
choices (\ref{paraval}) of $B^{1/4}$ and $\chi_0$ give reasonable results that
are
in the range of commonly-accepted phenomenology.
\bigskip

\noindent {\bf 5.2 Cluster distributions and hadron spectra}
\medskip

Using the parametrizations of Table 1, we have investigated
more quantitatively a number of typical features of the jet evolution,
which we discuss now.

In Fig. 9 we show the total transverse momentum generated during the
time evolution of the system in the center-of-mass of the initial
jet pair:
\begin{equation}
p^{(\alpha)}_\perp(t)\;\equiv\;
\int d^3 r \int dx dp^2 d p_\perp^2 \;p_\perp \;F_\alpha(t, \vec r;
x,p_\perp^2,p^2)
\label{pTtot}
\;,
\end{equation}
where $\alpha$ labels `partons' or `clusters', and
$p_\perp\equiv\sqrt{p_x^2+p_y^2}$.
As before, we compare the cases $Q=10 \;(100)$ GeV and
$L_c=0.6 \;(0.8)$ $fm$.
At $t=0$ we start with $p^{(\alpha)}_\perp(0)=0$, because the two initial
partons recede
back-to-back along the $z$ axis. Then, with progressing time,
the jet evolution can roughly be divided into four stages:
(i) a very short {\it hard stage}
($\,\lower3pt\hbox{$\buildrel < \over\sim$}\, 0.02$ $fm$),
characterized by an explosive production
of partons and consequently transverse momentum;
(ii) a longer {\it shower stage} ($\approx 0.02 - 0.3$ $fm$)
that essentially just causes diffusion in phase space
with little additional entropy and transverse momentum production;
(iii) a {\it conversion stage} ($\approx 0.3 - 5$ $fm$)
which sets in when the  partons start locally to form clusters;
(iv) a {\it hadronization stage}
($\,\lower3pt\hbox{$\buildrel > \over\sim$}\, 5$ $fm$)
when the clusters start fragmenting into physical hadron states
via cluster decay and resonances.
\smallskip

It is interesting to inspect
the distribution of the cluster sizes and the invariant-mass
spectrum of clusters, since these measures are essentially the only
microscopic information (aside from the momentum spectrum)
which is carried over from the partonic to the hadronic phase,
and therefore determines directly the final-state hadron distributions.
In Figs. 10 and 11  we show these distributions, for the two values of $L_c$
and the
two jet energies $Q$ considered before.

{}From Fig. 10a it is obvious that the typical cluster radius
is strongly peaked at a value slightly above $L_c$, with a very small
width of about $10^{-1}$ $fm$.
Specifically, the average cluster size is
$\langle R_{cl}\rangle = 0.62$ (0.83) $fm$ for $L_c=$ 0.6 (0.8) $fm$.
The narrow width of the cluster size distribution is a consequence of the
smallness of the surface tension $\sigma_c$ (\ref{sigval}), which implies a
very small surface energy to be overcome, and results in a jump
in the tunnelling probability (\ref{piV}) around $L=L_c$.
One also sees that there are a few clusters with radius smaller than $L_c$,
a feature which arises from the fact that cluster formation
sets in already at $L_\chi$ (c.f. Table 1), when
there is a non-zero transition probability for partons to yield clusters
by tunnelling through the potential barrier
that begins to develop when $L>L_\chi$ (c.f. Fig. 2).
A striking feature of the cluster mass spectrum shown in Fig. 10b
is that it is insensitive to the choice of $L_c$.
The spectrum is characterized by a strongly-damped exponential form at low mass
($M_{cl}\,\lower3pt\hbox{$\buildrel < \over\sim$}\, 3$ GeV),
with a high-mass tail that extends substantially beyond 5 GeV.
For both choices of $L_c$, the value for the average cluster mass is
$\langle M_{cl}\rangle \simeq 1.2$ GeV,
which is in qualitative agreement with the characteristic mass scale $m_\chi$
obtained in (\ref{chival}) as an estimate.

Figs. 11a and 11b display the corresponding size and mass distribution for the
case when we impose the constraint mentioned in Sec. 4.3, that two partons
cannot
form a cluster if their invariant mass is above a maximum
cluster mass $M_{crit}$, even if their mutual separation
increases beyond $L_c$, where we chose $M_{crit}=4$ GeV.
In this case the two partons propagate on as individual quanta until
they have either radiated off further energy, or recombine with
a less-energetic partner. Consequently, the cluster size distribution is
shifted to larger radii, not however very significantly,
because only a small fraction (typically less than 5 \%) of the
clusters are affected.
At the same time, the mass distribution loses its high-mass tail
and falls off considerably above 3 GeV,
and the average cluster mass comes down substantially to
$\langle M_{cl}\rangle \simeq 0.8$ GeV.

The most remarkable result of Figs. 10 and 11, however, is that the shapes of
both the
cluster size and cluster mass distribution are essentially independent of the
jet energy,
as well as of the initial 2-jet configuration, and therefore appear to be
universal.
\smallskip

The decay of the spectrum of formed clusters into hadrons, simulated according
to the Sec. 4.5,
then yields the average particle multiplicities of final-state hadrons.
It is interesting to look at the relation between parton and hadron
multiplicities.
The feature evident in Figs 10b and 11b, namely that
the mass spectrum of color-singlet clusters is independent of the total jet
energy,
is in agreement with analytical predictions \cite{MLLA1}.
This implies that parton and hadron
multiplicities should be proportional to each other at high energies,
which is known as {\it local parton-hadron duality} \cite{MLLA2}.
Fig. 12 display our results for
the total gluon and quark multiplcities
$\langle n_{qg} \rangle = \langle n_{g} \rangle + \sum_f \langle n_q+n_{\bar
q}\rangle$,
as well as
the ratios of charged hadrons to partons
$\langle n_{ch} \rangle /\langle n_{qg}\rangle$,
and of clusters to partons, $\langle n_{cl} \rangle /\langle n_{qg}\rangle$,
as a function of jet energy $Q$.
In Fig. 12a the calculated rise of
$\langle n_{qg}\rangle$ is shown for the two choices of $L_c$. The smaller
value of $L_c$ gives a larger multiplicity because we identified
$L_c^{-1}$ with the scale $\Lambda_{QCD}$ in $\alpha_s$, eq. (\ref{alpha}).
{}From Fig. 12b one reads off, however, that the average number of
clusters per parton
is independent of $L_c$ and of energy $Q$. As a consequence, the
number of charged hadrons per parton is larger for larger $L_c$,
because at any fixed $Q$ and due to 4-momentum conservation,
fewer partons yield more massive clusters which in turn
decay in a larger number of low-mass hadrons.
Most important, one sees that the ratio
$\langle n_{ch} \rangle /\langle n_{qg}\rangle$ is only weakly
energy-dependent for $Q\,\lower3pt\hbox{$\buildrel > \over\sim$}\, 30$ GeV,
in accord with the hypothesis of local parton-hadron duality:
it rises over this range by less than 10\% and appears to saturate
asymptotically,
approaching  a constant  of $\approx 1.6$ (1.7) for $L_c= 0.6$ (0.8) $fm$.

The resulting average multiplicity of charged hadrons as it rises with
$Q$ is shown in Fig. 13a together with experimental data \cite{pep,lep}.
In order to obtain the correct overall normalization, we
fitted the the infrared regulator $k_0$ entering $\alpha_s$, eq. (\ref{alpha}),
to
give the measured total charged multiplicity at $Q=91$ GeV for
$L_c=0.8$ $fm$. The required value is $k_0= 0.5$ GeV which implies $\alpha_s(0)
\approx 0.8$.
With this adjustment we then obtain in Fig. 13b the
momentum spectra of charged hadrons with respect to the variable
$\ln(1/x)$, where $x=2E/Q$ is the particle energy normalized to the
total energy $Q$.
The spectra clearly exhibit the well known `hump-back plateau'
\cite{MLLA2}.
The good agreement of the simulation with experimental spectra
is another indicator of the aforementioned
local parton-hadron duality. The result is not surprising, since
the simulation incorporates the {\it coherent} parton shower evolution
\cite{ms3}
based on the soft-gluon interference properties of the MLLA, which cause
this hump-back plateau with its depletion at small $x$.
Because in our approach cluster formation and subsequent
cluster decay involve no momentum dependence, but are  solely
described by the space-time separation of partons, the parton
momentum spectra in $x$ are  mapped almost unaltered onto the
hadron distributions.
We also conclude, that
the comparisons in Fig. 13 do not indicate any clear preference
for one value of $B$ or $L_c$ over the other.
However, as we will discuss in Sec. 5.3, an indication may be
drawn from Bose-Einstein correlations among produced hadrons.

An example of
the composition of the final hadron yield in terms of different
particle species is given in Table 2, where
we compare the result of our simulation of
$e^+e^-$ annihilation events at $Q=34$ GeV
with measured particle multiplicities  \cite{althoff83}.
The remarkable agreement is in accord with the presumption that
the different particle yields are essentially determined by
the available phase space and the density of hadron states,
and not by more complex mechanisms.
Further experimental constraints by more sensitive measures
of event shapes such as  thrust, sphericity, etc., may be investigated,
but it is evident from the results shown  that our approach
yields an overall satisfactory description that withstands  confrontation
with experiment, and encourages us to study more complex reactions
in the near future.
\bigskip

\noindent {\bf 5.3 The Bose-Einstein effect}
\medskip

{}From Figs. 10 and 11 we can conclude that the distribution
of formed clusters clearly resembles the picture of
{\it preconfinement} \cite{preconf}, which is the tendency of partons
produced during the cascade evolution
to arrange themselves in color-singlet clusters with limited
extension in both position and momentum space.
Since the clusters
are the basic units within which the final-state hadrons
arise, the ensemble of clusters in phase space, as it
builds up with time, can be interpreted
as a particle emission source with a space-time distribution
that is determined by the preceding parton evolution.
This notion allows us to directly relate the dynamics of
cluster formation to the well-known Bose-Einstein effect \cite{bowler85},
which corresponds to an enhancement in the production rate
of identical bosons (in our case mainly pions) emitted from
similar regions in space and time, arising from the imposition
of Bose symmetry.
Enhancements in the mass spectrum of {\it same-sign} pion pairs have been
seen clearly in $e^+e^-$-data (for a review see e.g. \cite{haywood94}).
Let us briefly recall that the Bose symmetry imposed on the production
amplitude of identical particles from a distribution of sources
leads to an interference term in
the squared amplitude which is only observable if the sources are
incoherent.
{}From the analysis of $e^+e^-$-data one finds \cite{haywood94} that the
Bose-Einstein
effect is reasonably described by a spherically-symmetric
space-time distribution of sources with Gaussian form
\begin{equation}
\rho(r)\;=\; \rho(0)\;\exp\left(-\frac{r^2}{2 \sigma_\rho^2}\right)
\;,
\end{equation}
where $\sigma_\rho$ is a radius parameter. Such a source leads to an
enhancement
due to interference caused by the identical-particle effect,
relative to the rate with no interference,
\begin{equation}
b(q)\;=\; 1\;\,+\;\, \lambda_\rho
\;\exp\left(-\sigma_\rho^2\,q^2\right)
\;,
\label{bk}
\end{equation}
where $q^2 = m_{\pi\pi}^2 - 4 m_\pi^2$, and $m_{\pi\pi}=\sqrt{(p_1+p_2)^2}$ is
the
invariant mass of the emitted pion pair. The degree of incoherence of the
source is measured by $\lambda_\rho$ (=1 for complete incoherence, and =0 for
complete coherence), $\sigma_\rho$ measures the source size in $fm$, or
alternatively,
$\sigma_\rho^{-1}$ measures the range of enhancement with respect to $q$ in
GeV.

To observe an enhancement in $q$ amongst identical particles,
one must compare the particle distributions
with corresponding spectra in the complete absence of Bose
symmetry.
Thus, in order to get an estimate of the magnitude of enhancement
implied by our hadronization picture, we proceed as follows.
First we evaluate the pion distributions resulting from a simulation
which does not include the Bose-Einstein effect. Then we repeat
the calculation, but now imposing Bose symmetrization on
same-sign pion pairs by assuming complete incoherence, corresponding to
$\lambda_\rho=1$. Here we use the method of
Sj\"ostrand \cite{sjo93}, which simulates the enhancement
due to the identical-particle effect.
Finally, we compute the ratio $b(q)$
of the pion distributions of same-sign pairs with Bose symmetrization
to the one without, as a function of the invariant mass $q$.
In Fig. 14a we show the resulting enhancement $b_{L_c}(q)$
for our two previously-used values of $L_c$, confronted with the
corresponding distribution obtained by the OPAL collaboration \cite{BE} at
$Q=91$ GeV.
It is remarkable how well this comparison with the experimental data
allows us to separate the two choices of $L_c$ in our calculations.
Clearly the value $L_c=0.8$ $fm$ appears to be strongly favored.
In fact, the average source size in this case turns out
$\sigma_\rho = 0.84$ $fm$, which is almost identical
to the average cluster size determined from Figs. 10 and 11.
On the other hand, this value is well in the range
of the pion source radius determined by OPAL, $\sigma_\rho^{exp}= 0.93 \pm
0.17$ $fm$
with $\lambda_\rho^{exp} = 0.87 \pm 0.14$.

We may thus conclude  that our presumed identification of
$L_c\simeq \sigma_\rho$ indeed has physical relevance that provides a
unique relation between the parameter $L_c$ and the experimentally-observed
pion emission source radius.  With this important insight, it would be
interesting
to investigate this issue in more detail,
because it provides a promising method to extract
properties of the partons' space-time evolution
and cluster formation from the measured particle distributions.
Since the structure of the perturbative
parton cascade development is projected locally onto the cluster distribution,
which itself maps on the hadron spectra, the characteristic shape
of the Bose enhancement $b(q)$ will depend only on the local environment,
which may in turn depend on the physical situation
(vacuum, as considered here, or medium, as, e.g., in deep-inelastic
lepton-nucleus scattering or nucleus-nucleus collisions).
Thus, by comparing, for instance,
the Bose-Einstein correlations measured in $e^+e^-\rightarrow hadrons$
to high-energy heavy-ion collisions, one might
extract specific features of perturbative QCD in a finite-density and
-temperature
medium, which are absent in vacuum \cite{msrep,mspa}.

As an illustrative example of such a comparison, we show in Fig. 14b
the ratio  $b_{0.6\,fm}(q)/b_{0.8\,fm}(q)$ of the
curves in Fig. 14a.
Although the individual curves in Fig. 14a are very similar to each other,
their ratio is a very sensitive quantity that filters out clearly  their subtle
difference.
It is evident that a smaller $L_c$ gives
rise to a significantly stronger enhancement at
low masses
$k\,\lower3pt\hbox{$\buildrel < \over\sim$}\, 300$ MeV,
peaked at about 1.5 times the pion mass.
Fig. 14b also shows that the results for $Q=34$ GeV and $Q=91$ GeV are
identical, even for this sensitive ratio, which implies that the
specifics of the Bose-Einstein effect and Bose enhancement are independent of
the
energy, in agreement with what is observed experimentally \cite{haywood94}.
\bigskip

\bigskip
\noindent {\bf 6. SUMMARY AND PERSPECTIVES}
\bigskip

In conclusion, we have presented a novel approach to the
dynamics of parton-hadron conversion and confinement, based on
an effective QCD field theory and a kinetic multi-particle description
in real time and complete phase space.
Our formulation provides an extension of the well-understood perturbative QCD
parton evolution to account for the full space-time history
traced from parton cascade development, via cluster formation and decay, all
the way
to the production of final hadrons.
The essential points in our approach may be summarized as follows.
\smallskip

\noindent {\bf (i)}
We have constructed a scale-dependent Lagrangian that incorporates
{\it both} parton and hadron degrees of freedom. It is manifestly
gauge- and Lorentz-invariant, and consistent with the scale and
chiral symmetry properties of QCD.
The introduction of the scale $L(r)$ determines
locally which are the relevant degrees of freedom around a given
space-time point $r$.
\smallskip

\noindent {\bf (ii)}
The formulation
recovers  QCD with its symmetry properties at short space-time distances, and
merges into an effective low-energy description of hadronic degrees of freedom
at large
distances. In between the two regimes it interpolates as determined
by the scale- ($L$-)changing dynamics, and results in a transformation from
partonic to hadronic
degrees of freedom.
\smallskip

\noindent {\bf (iii)}
The dynamics is described by a set of coupled kinetic equations that derive
from the field equations of motion, and yield a real-time description in
both position and momentum space, constrained by the uncertainty principle.
\medskip

As a test application, we have considered the prototype reaction
$e^+e^- \rightarrow hadrons$ where the fragmentation of
parton jets and their hadronization serves as a generic
process that can also be imagined as an integral part of
more complex reactions.
We investigated in detail the specifics of the time evolution of
parton shower, cluster formation
and hadron production in phase space, which extends the usual
QCD evolution techniques that are limited to momentum space and integrated over
time.
The consistency with experimental data was tested, and we found good
agreement with measured hadron spectra. A prospective method
to extract the characteristics of the space-time development
from Bose-Einstein correlations of identical hadrons was suggested.
Our main results are:
\smallskip

\noindent {\bf (i)}
The details of the parton-hadron conversion are controlled by
the quantity $L_c$,
the spatial separation of neighbouring color charges in their restframe,
which defines the scale at which non-perturbative confinement forces become
substantial.
The value of $L_c$ is determined by the choice of the bag constant $B$ and the
condensate value $\chi_0$.
All other macroscopic quantities which
characterize the evolution, such as pressure, energy density,
the time scale of the transition, etc., are
then determined self-consistently. The found values are in agreement
with common phenomenology.
\smallskip

\noindent {\bf (ii)}
The time scale obtained for the transition is a remarkable result:
the local conversion of partons to clusters occurs very rapidly
($\approx 0.1-0.2$ $fm$), but the global time scale for the transition of the
system as a whole is long ($\approx 10-30$ $fm$).
\smallskip

\noindent {\bf (iii)}
The QCD features of the perturbative parton evolution are projected
unscathed onto cluster and hadron distributions,
because the conversion is a local, universal mechanism.
As a consequence the multiplicities of charged hadrons and their momentum
spectra
are predetermined by the preceding parton evolution, which we find in good
agreement with experiments.
\smallskip

\noindent {\bf (iv)}
Our main result is the sensitivity of the Bose-Einstein
correlations among identical pions due to Bose symmetry.
It allows us to identify the parameter $L_c$ with the hadron emission source
radius measured in experiments, and to fix its value rather precisely to
$L_c\simeq 0.8$ $fm$.
Moreover, the ratio of Bose enhancement of same-sign pions
in different scenarios or physical situations can provide
a very sensitive probe of the environment in which the parton system
evolves. It can be exploited to study, e.g., modifications
to parton evolution in finite-density media.
\medskip

\noindent
It is interesting to note that our model appears to be able
to correlate sucessfully such diverse quantities as the
macroscopic bag constant $B$,
the microscopic length scale $L_c$ for the
parton-hadron transition, the associated critical temperature $T_c$,
and the magnitude of measured Bose-Einstein correlations.
\smallskip

Finally we comment on future applications. These are manifold,
as the advocated picture of parton-hadron conversion is universally
applicable to any dynamical process where the issue of hadronization arises.
Since the strength of our statistical real-time description lies in
resolving the details of the space-time structure, situations where
parton cascades undergo interactions with an environment would be the
interesting to investigate. Let us give three examples:
\smallskip

\noindent {\bf a)}
In {\it deep-inelastic lepton-nucleus scattering}
the primary quark struck by the photon can travel and reinteract before
hadronizing, and produce a cascade of secondary partons that
differs from a parton shower in vacuum. The secondaries are
themselves potential candidates
for hard re-interactions, and can lead to a specific $A$ (atomic number)
dependence for the final-state hadron production.
Clearly, here it is essential to keep track of the parton-hadron conversion
at each point in time and space, because partons that reinteract
will not be able to hadronize before they approach the free-streaming regime.
\smallskip

\noindent {\bf b)}
In {\it high-energy nucleus-nucleus collisions} \cite{qm93}
($\sqrt{s}/A\,\lower3pt\hbox{$\buildrel > \over\sim$}\, 200$ GeV)
the parton density of the highly Lorentz-contracted nuclei
is very large already in the initial state, and is further increased
by the materialization and multiplication of partons \cite{msrep}.
Therefore multiple scatterings
of partons can easily lead to a large number of
simultanously-evolving cascades that also can interact with each other.
In order to resolve such an intertwined structure of parton interactions,
the space-time dynamics of the system {\it must} be taken into account.
Again, here a microscopic space-time description of parton-hadron conversion
is crucial to resolve the details of
such an intertwined structure of parton interactions and the
following hadron formation process, depending on the local
densities of surrounding parton and hadron matter.
\smallskip

\noindent {\bf c)}
The {\it QCD phase transition} from a hot, deconfined quark-gluon plasma
to excited hadron matter as occurred in the early Universe \cite{ellis93}
is of long-standing theoretical interest. Lattice QCD
calculations to date can only investigate the critical behaviour
in the vicinity of the transition temperature.
Moreover, a dynamical evolution of the system deviating
significantly from thermal equilibrium is not achievable.
In view of the future experimental programs at RHIC and LHC,
it will soon become possible
to recreate the QCD phase transition in the laboratory \cite{bm93},
and to investigate  its dynamics in the real world.
It is clear that the conversion of a quark-gluon plasma into hadrons
is a much more complicated process than the hadronization of final-state
partons in free space (as in $e^+e^-$ annihilation),
or dilute systems (as in hadron-hadron collisions).
In heavy-ion collisions the transition is expected to
proceed as a complex evolution of expansion and cooling of the
plasma, perhaps through a mixed parton-hadron phase, until
a purely hadronic phase is reached \cite{csernai93}.
Therefore, a fully dynamical and microscopic hadronization scheme
as proposed in this paper is needed to trace the space-time-dependent
cooling and expansion process from the parton to the hadron phase.
\smallskip

%\newpage

\newpage

{\bf TABLE CAPTIONS}
\bigskip

\noindent {\bf Table 1:}
Obtained values for  phenomenological quantities involved in our description.
The bag constant $B$ and the condensate value
$\chi_0=\langle 0|\chi|0\rangle$ are fixed inputs.
The values of the length scales $L_c$, $L_\chi$, the global conversion
time scale $\tau_0\propto L_0$, and the magnitude of
the surface tension $\sigma_c$, are then extracted from the numerical
simulation.
Also listed is the associated critical temperature $T_c$, the
mass scale of the lightest scalar glueball $m_\chi$, and the
estimate for the gluon condensate $G_0$.
\bigskip

\noindent {\bf Table 2:}
$e^+e^-\rightarrow hadrons$ at $Q=34$ GeV:
Average multiplicities of partons
$\langle n_{qg} \rangle = \langle n_{g} \rangle + \sum_f \langle n_q+n_{\bar
q}\rangle$,
of clusters $\langle n_{cl} \rangle$, and of charged hadrons
$\langle n_{ch} \rangle$, plus the
contribution of pions, kaons and protons,
in comparison  with measured particle multiplicities  \cite{althoff83}.

\bigskip
\bigskip

{\bf FIGURE CAPTIONS}
\bigskip

\noindent {\bf Figure 1:}
Schematic behaviour of $\kappa_L(\chi)$, eq. (\ref{kappa}) and $\mu(\chi)$, eq.
(\ref{Mchi}).
\bigskip

\noindent {\bf Figure 2:}
Form of the scale-dependent potential ${\cal V}(L)$, eq. (\ref{calV}),
where $L_\chi$ characterizes the point of inflection at $\chi_\chi$,
$L_c$ marks the point when the two minima are degenerate at $\chi_c$,
and $L_0$ when the potential has a single absolute minimum at $\chi_0$.
The value at $\chi=0$ is equal to the vacuum pressure (bag constant) $B$.
\bigskip

\noindent {\bf Figure 3:}
Diagrammatic representation of
{\bf a)}
the two-point Green functions $S, D, \Delta, \tilde \Delta$;
{\bf b)}
the self energies $\Sigma, \Pi, \Xi, \tilde \Xi$;
{\bf c)}
the corresponding Dyson-Schwinger equations (\ref{sde}).
{\it Notation}:
Fat (thin) lines indicate fully-dressed (bare) propagators,
shaded circles and boxes denote the full quark and gluon vertex functions,
black circles and boxes with attached loops represent
the local interactions with the collective fields $\chi$ and $U$
via the potential ${\cal V}$, eq. (\ref{calV}).
\bigskip

\noindent {\bf Figure 4:}
Illustration of the simultanous evolution in
space-time $r$ and $-$ `orthogonal' to it $-$ in energy-momentum $p$
of the Wigner functions $F_\alpha(r,p)$, according to
the transport and constraint equations (\ref{T}) and (\ref{R}).
The self-consistent solution of these equations corresponds to summing over
all possible quantum paths $r$, accounting for fluctuations in $p$,
under the constraint of the uncertainty principle.
\bigskip

\noindent {\bf Figure 5:}
Schematics of the different stages of the process
$e^+e^-\rightarrow hadrons$. The initial $q\bar q$ pair
with large invariant mass $Q$ initiates a shower which can be followed
perturbatively until
$L\simeq L_\chi$. At this point the conversion into clusters sets in,
which is completed around $L=L_0$ and followed by the decay of clusters
into hadron states.
\bigskip

\noindent {\bf Figure 6:}
Diagrammatic representation of the kinetic equations (\ref{eu}).
{\bf a)}
The operator $\hat{\cal K}$ describes free propagation plus
the effect of the mean field.
{\bf b)}
The integral operators $\hat A, \hat B, \hat C, \ldots$
include the squared amplitudes for the various interaction
processes among the different particle species, which
change of the particle distributions
according to the balance of gain (+) and loss ($-$) terms.
\bigskip

\noindent {\bf Figure 7:}
{\bf a)}
Space-time evolution of the parton density profile (in arbitrary units)
in the ($r_z,r_\perp$)-plane at different times in the center-of-mass
frame of the initial dijet system  with energy $Q=100$ GeV.
{\bf b)}
Corresponding development of the cluster density profile
as it builds up in time due to the conversion of partons.
\bigskip

\noindent {\bf Figure 8:}
{\bf a)}
Time evolution of the kinetic pressures $P_{qg}$ of partons and $P_\chi$
of pre-hadronic clusters for $q\bar q$-initiated jet evolution,
for total jet energies  $Q=10$ GeV (top) and  $Q=100$ GeV (bottom).
The dashed and full lines correspond to
the two parameter choices $(B^{1/4},\chi_0)=$(240,200) MeV
and $(B^{1/4},\chi_0)=$(180,100) MeV, resulting in
$L_c=0.6$  $fm$ and $L_c=0.8$ $fm$, respectively.
{\bf b)}
As a), but for a $gg$-initiated jet evolution.
\bigskip

\noindent {\bf Figure 9:}
Total transverse momentum $p_\perp(t)$, eq. (\ref{pTtot}), generated during the
time evolution of the system in the center-of-mass of the initial
dijet system,  in correspondence to Fig. 8:
{\bf a)}
case of  $q\bar q$-initiated jet evolution (top);
{\bf b)}
case of a $gg$-initiated jet evolution (bottom).
\bigskip

\noindent {\bf Figure 10:}
Cluster spectra for $L_c=0.6$ $fm$ (top), and $L_c=0.8$ $fm$ (bottom),
and total jet energies $Q=10$ (100) GeV.
{\bf a)}
Distribution of the cluster sizes of clusters formed from neighboring partons.
{\bf b)}
Associated cluster mass spectrum.
\bigskip

\noindent {\bf Figure 11:}
As Fig. 10, but now with the additional constraint of a maximum allowed
invariant mass per cluster of $M_{crit}=4$ GeV:
{\bf a)}
cluster size distribution;
{\bf b)}
cluster mass spectrum.
\bigskip

\noindent {\bf Figure 12:}
{\bf a)}
Total gluon and quark multiplcities
$\langle n_{qg} \rangle = \langle n_{g} \rangle + \sum_f \langle n_q+n_{\bar
q}\rangle$,
for $L_c=0.6$ (0.8) $fm$ as a function of energy $Q$.
{\bf b)}
The corresponding ratios of charged hadrons to partons
$\langle n_{ch} \rangle /\langle n_{qg}\rangle$,
and of clusters to partons, $\langle n_{cl} \rangle /\langle n_{qg}\rangle$.
\bigskip

\noindent {\bf Figure 13:}
{\bf a)}
Calculated average charged multiplicity versus total energy
$Q$ in $e^+e^-$ annihilation events, in comparison with experimental data
\cite{pep}.
{\bf b)}
Momentum spectra of charged hadrons with respect to the variable
$\ln(1/x)$, where $x=2E/Q$, at $Q=34$ GeV and $Q=91$ GeV,
confronted with  distributions measured
at PEP/PETRA and LEP \cite{lep}.
\bigskip

\noindent {\bf Figure 14:}
{\bf a)}
Simulated Bose-Einstein enhancement $b_{L_c}(q)$ as a
function of the pair mass $q$ of same-sign pion pairs
for the two values of $L_c$ at total energy
$Q=91$ GeV. The data points are from the OPAL experiment \cite{BE}
at LEP.
{\bf b)}
Ratios of the enhancements $b_{0.6\,fm}(q)/b_{0.8\,fm}(q)$
for total jet energies $Q=34$ GeV and $Q=91$ GeV.

\newpage

\begin{center}
{\normalsize

\begin{tabular}{c|ccc}
\hline
\hline
%     &    &&      \\
          &  $\;\;\;\;\;\;\;$   $B^{1/4} \;=\; 230$ MeV   &$\;\;$&
$B^{1/4}\;=\; 180$ MeV \\
          &  $\;\;\;\;\;\;\;$   $\chi_0 \;=\; 200$ MeV   &$\;\;$&
$\chi_0\;=\; 100$ MeV \\
%     &    &&      \\
\hline
\hline
$\;\;\;\;\;\;\;\;\;\;\;\;\;\;\;\;\;$ &&    $\;\;\;\;\;\;\;\;\;\;\;\;\;\;\;\;\;$
  &    \\
%            &            &&                       \\
$L_c$ (fm)    $\;\;\;\;\;$&   0.6  &&  0.8 \\
%            &            &&                       \\
$L_\chi$ (fm) $\;\;\;\;\;$&   0.45  &&  0.65 \\
%            &            &&                       \\
$\tau_0$ (fm) $\;\;\;\;\;$&   8.5 (21)  &&  10 (26) \\
            &            &&                       \\
$\sigma_c^{1/3}$ (MeV) $\;\;\;\;\;$ &   40  &&  48 \\
%            &            &&                       \\
$T_c$ (MeV) $\;\;\;\;\;$ &  160  &&  125 \\
%            &            &&                       \\
$m_\chi$ (GeV) $\;\;\;\;\;$ &  1.05  &&  1.30 \\
%            &            &&                       \\
$G_0$ (GeV/fm$^3$) $\;\;\;\;\;$ &  1.25  &&  0.50 \\
            &            &&                       \\
\hline
\hline
\end{tabular}

}

\bigskip

{\Large {\bf Table 1}}
\end{center}

\bigskip \bigskip
\bigskip

\begin{center}

\begin{tabular}{l|llllllllc}
\hline
\hline
                                            &&
&$\;\;\;\;\;\;\;\;\;\;\;\;$&      &$\;\;\;\;\;\;\;\;\;\;\;\;$&      &&&
       \\
$Q = $ 34 GeV                             &&   & $L_c=$ 0.6 fm &   & $L_c=$ 0.8
fm  &  &   &&  Experiment \hspace{0.3cm}\\
                                            &&      &      &      &      &
&      && Ref. \cite{althoff83}               \\
\hline
\hline
                                            &&      &      &      &      &
&      &&                \\
$\langle n_{qg} \rangle$                    &&  & 9.7 &  & 8.6 &  &  &&   $-$
\\
$\langle n_{cl} \rangle$                    &&  & 8.7 &  & 7.7 &  &  &&   $-$
\\
%                                           &&      &      &      &      &
%%&      &&                \\
$\langle n_{ch} \rangle$                    &&  & 14.1 &  & 13.5 &  &  && 13.6
$\pm$ 0.9 \\
$\langle n_{\pi^+}+n_{\pi^-} \rangle$       &&  & 11.4 &  & 10.9 &  &  && 10.3
$\pm$ 0.4 \\
$\langle n_{K^+}+n_{K^-} \rangle$           &&  &  1.6 &  &  1.5 &  &  &&  2.0
$\pm$ 0.2 \\
$\langle n_{p}+n_{\bar p} \rangle$          &&  &  0.8 &  &  0.7 &  &  &&  0.8
$\pm$ 0.1 \\
                                            &&      &     &      &      &
&      &&                \\
\hline
\hline
\end{tabular}

\bigskip

{\Large {\bf Table 2}}
\end{center}

\vfill

\end{document}